\documentclass[aps, showpacs,preprintnumbers, superscriptaddress, nofootinbibt, twocolumn]{revtex4-1}
\usepackage{makeidx}
\usepackage{amsfonts}
\usepackage{amsmath}
\usepackage{amssymb}
\usepackage{eurosym}
\usepackage{graphicx}

\def\be{\begin{equation}}
\def\ee{\end{equation}}
\def\bea{\begin{eqnarray}}
\def\eea{\end{eqnarray}}

\def\nn{\nonumber \\}

\usepackage{color}

\begin{document}

\title{Coupling matter and curvature in Weyl geometry: conformally invariant $f\left(R,L_m\right)$ gravity}
\author{Tiberiu Harko}
\email{tiberiu.harko@aira.astro.ro}
\affiliation{Department of Theoretical Physics, National Institute of Physics
and Nuclear Engineering (IFIN-HH), Bucharest, 077125 Romania}
\affiliation{Astronomical Observatory, 19 Ciresilor Street,
	Cluj-Napoca 400487, Romania,}
\affiliation{Department of Physics, Babes-Bolyai University, Kogalniceanu Street,
	Cluj-Napoca 400084, Romania,}
\affiliation{School of Physics, Sun Yat-Sen University, Guangzhou 510275, People's
	Republic of China,}
\author{Shahab Shahidi}
\email{s.shahidi@du.ac.ir}
\affiliation{School of Physics, Damghan University, Damghan, 41167-36716, Iran}

\date{\today }

\begin{abstract}
We investigate the coupling of matter to geometry in conformal quadratic Weyl gravity, by assuming a coupling term of the form $L_m\tilde{R}^2$, where $L_m$ is the ordinary matter Lagrangian, and $\tilde{R}$ is the Weyl scalar. The coupling explicitly satisfies the conformal invariance of the theory. By expressing $\tilde{R}^2$ with the help of an auxiliary scalar field and of the Weyl scalar, the gravitational action can be linearized, leading in the Riemann space to a conformally invariant $f\left(R,L_m\right)$ type theory, with the matter Lagrangian nonminimally coupled to the Ricci scalar. We obtain the gravitational field equations of the theory, as well as the energy-momentum balance equations. The divergence of the matter energy-momentum tensor does not vanish, and an extra force, depending on the Weyl vector, and matter Lagrangian is generated. The thermodynamic interpretation of the theory is also discussed. The generalized Poisson equation is derived, and the Newtonian limit of the equations of motion is considered in detail. The perihelion precession of a planet in the presence of an extra force is also considered, and constraints on the magnitude of the Weyl vector in the Solar System are obtained from the observational data of Mercury. The cosmological implications of the theory are also considered for the case of a flat, homogeneous and isotropic Friedmann-Lemaitre-Robertson-Walker geometry, and it is shown that the model can give a good description of the observational data for the Hubble function up to a redshift of the order of $z\approx 3$.
\end{abstract}

\maketitle
\tableofcontents

\section{Introduction}

The birth of general relativity through the work by Einstein \cite{Eina} and Hilbert \cite{Hil} did have a deep impact not only on physics, but also on mathematics, leading to several extensions of the Riemannian geometry. Almost immediately after general relativity was proposed, Weyl \cite{Weyl} did develop a generalization of Riemann geometry, with the explicit goal of formulating a unified theory of gravity and electromagnetism. The starting point in Weyl's approach was the observation that in vacuum  Maxwell's equations are conformally invariant, which led him to suggest that the gravitational field equations should have the same symmetry. In Weyl's geometry the covariant derivative of the metric tensor is conjectured to be nonzero, so that $\nabla _{\lambda}g_{\mu \nu}=Q_{\lambda \mu \nu}=\omega_{\lambda}g_{\mu \nu}$, where the new geometric quantity $Q_{\lambda \mu \nu}$ is called the nonmetricity, while $\omega _{\lambda}$ is the Weyl vector field. For a detailed account of Weyl theory and its possible physical interpretation see \cite{Weyl1}. Moreover, in a Weyl geometric framework the parallel transport does not keep the length of a vector constant. This feature of the Weyl geometry led to Einstein's severe criticism of its initially proposed physical interpretation, based on the claim that since the behavior of the atomic clocks would depend on their past history, the existence of sharp spectral lines in the presence of an electromagnetic field would be impossible.

If one gives up the interpretation of the Weyl vector field as an electromagnetic type potential, the conformally invariant Weyl geometry represents an important and beautiful  generalization of Riemannian geometry. In the following {\it we consider the Weyl field as having a purely geometric nature}. The physical investigations using Weyl geometry are built on the fundamental assumption {\it that conformal invariance is a basic symmetry property of Nature}. A local conformal transformation
\begin{equation*}
d\tilde{s}^2=\Sigma ^2(x)ds^2=\Sigma ^2(x)g_{\mu \nu}dx^{\mu}dx^{\nu}=\tilde{g}_{\mu \nu}dx^{\mu}dx^{\nu},
\end{equation*}
 {\it does not transform the local coordinates}, but {\it changes the units for the measurements in space-time}. The Christoffel symbols transform under a conformal transformation according to \cite{Scholz}
 \begin{equation*}
 \tilde{\Gamma}^{\lambda}_{\mu \nu}=\Gamma _{\mu \nu}^{\lambda}+\Big[\frac{\Sigma_{,\mu}}{\Sigma}\delta_{\nu}^{\lambda}+\frac{\Sigma _{,\nu}}{\Sigma}\delta_{\mu}^{\lambda}-g^{\lambda \sigma}\frac{\Sigma_{,\sigma}}{\Sigma}g_{\mu \nu}\Big].
 \end{equation*}

 If $\tilde{\omega}_{\mu}=\omega _{\mu}+2\Sigma _{,\mu}/\Sigma$, then $\tilde{\Gamma}^{\lambda}_{\mu \nu}=\Gamma _{\mu \nu}^{\lambda}$, $\tilde{R}^{\mu}_{\nu \lambda \sigma}=R^{\mu}_{\nu \lambda \sigma}$, $\tilde{R}_{\mu \nu}=R_{\mu \nu}$, and $\tilde{F}_{\mu \nu}=F_{\mu \nu}$, where $F_{\mu \nu}=\nabla _{\mu}\omega_{\nu}-\nabla _{\nu}\omega _{\mu}$.
Hence, one can consider $\omega _{\mu}$ as {\it a gauge field mediating at different space-time points the conformal factors} \cite{Bera, Berb}.

 Weyl gravity was generalized by Dirac \cite{Dirac1,Dirac2}, who introduced in the theory a real scalar field $\beta$ of weight $w(\beta)=-1$.  The corresponding gravitational Lagrangian is given by
 \be\label{LD}
 L=-\beta ^2R+kD^{\mu}\beta D_{\mu}\beta +c\beta ^4+\frac{1}{4}F_{\mu \nu}F^{\mu \nu},
 \ee
  where $k=6$ is a constant. The Lagrangian given by Eq.~(\ref{LD}) has the important property of conformal invariance. The cosmological implications of the  Lagrangian (\ref{LD}) were investigated in \cite{Rosen}. An alternative Weyl-Dirac type Lagrangian was suggested in \cite{Isrcosm}, and it is given by
 \bea
 L&=&W^{\lambda \rho}W_{\lambda \rho}-\beta ^2R+\sigma \beta ^2w^{\lambda}w_{\lambda}+2\sigma \beta w^{\lambda}\beta _{,\lambda}+\nonumber\\
&& (\sigma +6)\beta _{,\rho}\beta_{,\lambda }g^{\rho \lambda}+2\Lambda \beta ^4+L_m,
 \eea
 where $W_{\mu \nu}$ is the Weyl curvature tensor, constructed with the help of the  Weyl vector $\omega_{\mu}$, and $\beta $ is the Dirac scalar field, respectively. $\sigma$ and $\Lambda$ are constants. An interesting property of this model is that ordinary matter is created at the beginning
 of the cosmological evolution, due to the presence of the Dirac gauge function. Moreover, in the late Universe,  Dirac’s gauge function creates the dark energy that determines the recent acceleration of the Universe. For other physical generalizations of Weyl theory see \cite{Ut1,Ut2,Ni}.

 Independently of Weyl geometry, but inspired by it, {\it the idea of the conformal invariance of the gravitational action in Riemann geometry} has attracted a lot of attention. Gravitational theories based on the action
\begin{equation}\label{W1}
S_{Weyl} =-\frac{1}{4}\int d^4x\sqrt{-g}
C_{\mu\nu\rho\sigma}C^{\mu\nu\rho\sigma},
\end{equation}
where $C_{\mu\nu\rho\sigma}$ is {\it the Weyl tensor}, are called {\it conformally invariant, or Weyl
gravity type theories}. They have been investigated in detail in \cite{Ma0, Ma1,Ma2,Ma3,Ma4,Ma5,Ma6,Ma7,Ma8,Ma9,Ma10,Ma11,Ma12,Ma13,Ma14,Ma15,Ma16}. In particular, for a  static spherically symmetric metric of the form $ds^2=-B(r)dt^2+B^{-1}(r)dr^2+r^2d\Omega$, the theory admits vacuum solutions of the form $B(r)=1-3\beta \gamma -\beta\left(2-3\beta \gamma\right)/r+\gamma r+kr^2$, where $\beta$, $\gamma$ and $k$ are constants \cite{Ma0}. It was also suggested that Weyl gravity could explain the flat rotation curves of galaxies without the need of introducing dark matter \cite{Ma0}.

An important application of Weyl geometry to the study of gravitational interaction is the $f(Q)$ gravity theory, or the symmetric teleparallel gravity. Initially proposed in \cite{Q1}, in this theory the nonmetricity $Q$ of a Weyl geometry is the fundamental geometrical quantity describing all the physical aspects of gravity. This approach to gravity was extended in \cite{Q2}, and is known presently as the $f(Q)$ gravity theory, or nonmetric gravity \cite{Q2}. In the presence of matter the action of the gravitational field is given by $S=\int{\left(f(Q)+L_m\right)\sqrt{-g}d^4x}$. The physical, cosmological and geometrical properties of the $f(Q)$ theory have been studied  in detail in \cite{Q4,Q5,Q6,Q7,Q8,Q9,Q10,Q11,Q12,Q13,Q14,Q15,Q16,Q17, Q18, Q19,Q20,Q21,Q22,Q23}.

The role of the conformal structures in cosmology was emphasized by Penrose \cite{P1}, who, based on the observation that at the end of the de Sitter accelerating expansionary phase, induced by the presence of the positive cosmological constant $\Lambda$, the spacetime will be space-like, and conformally flat, as it was the initial boundary of the Universe during the Big Bang, introduced a cosmological scenario called {\it Conformal Cyclic Cosmology} (CCC). In this model the Universe consists of eons, representing a time oriented spacetime, with their conformal compactifications having spacelike
null infinities. Different aspects of the  Conformal Cyclic Cosmology were investigated in \cite{P1a, P2,P3,P4,P5,P6,P7}. In particular, in \cite{P2} a Weyl-invariant action that describes both gravity and the standard model, with Lagrangian
\bea
L&=&\frac{1}{12}\left( \phi ^{2}-2H^{\dagger }H\right) R\left( g\right)+g^{\mu \nu }\Bigg( \frac{1}{2}\partial _{\mu }\phi \partial _{\nu }\phi \nonumber\\
&&
-D_{\mu }H^{\dagger }D_{\nu }H\Bigg) -\left[ \frac{\lambda }{4}\left(
H^{\dagger }H-\omega ^{2}\phi ^{2}\right) ^{2}+\frac{\lambda ^{\prime }}{4}%
\phi ^{4}\right] \nonumber\\
&&+L_{SM},
\eea
was investigated, where $L_{SM}$ denotes the standard model Lagrangian, without the
kinetic and self interaction terms of the Higgs doublet $H$, the scalar
field $\phi $ is a singlet under $SU(2)\times U(1)$, which does not couple
to the standard model fields, with the exception of the Higgs field, while $\omega $ is a
small parameter, of the order of $10^{-17}$, determining the Higgs vacuum
expectation value, and the Higgs mass.

The importance of the local conformal symmetry was emphasized by 't Hooft in \cite{G1a}, where it was argued that it is {\it an exact symmetry that is spontaneously broken}. The breaking of the conformal symmetry may lead to a mechanism unveiling the small-scale structure of the gravitational interaction. {\it This symmetry could be as important as the Lorentz invariance}, and could help in the understanding of the Planck scale physics. A theory of gravity based on the idea that local conformal symmetry is an exact, but spontaneously broken symmetry of nature was proposed in \cite{G2}. The Lagrangian of the theory is $L=L_{EM}+L_{matter}$, where
\begin{equation}
L_{EM}=\sqrt{-\hat{g}}\left[ \frac{1}{16\pi G}\left( \omega ^{2}\hat{R}+6%
\hat{g}^{\mu \nu }\partial _{\mu }\omega \partial _{\nu }\omega \right) -%
\frac{\Lambda }{8\pi G}\omega ^{4}\right] ,
\end{equation}%
and
\bea
L_{matter}&=&-\frac{1}{4}F_{\mu \nu }F^{\mu \nu }+\sqrt{-\hat{g}}\Bigg( -\frac{%
1}{2}6\hat{g}^{\mu \nu }D_{\mu }\phi D_{\nu }\phi \nonumber\\
&&-\frac{1}{2}m^{2}\omega\overleftarrow{}
^{2}\phi ^{2}-\frac{1}{2}\phi ^{2}\hat{R}-\frac{\lambda }{8}\phi ^{4}\Bigg)
+L_{ferm},
\eea
where the term  $\phi^2R^2$ was included for restoring the conformal invariance of $L_{matter}$, and $\hat{R}$  is the scalar curvature constructed  from $g_{\mu \nu}$. In this model the conformal component of the metric field can be treated as a dilaton field, and a black hole turns into a  topologically trivial, regular soliton without horizons, firewalls, and singularities.

Conformal Weyl gravity, quadratic in the scalar curvature, and in the Weyl tensor, was investigated, in both metric and Palatini formulations, in \cite{Gh1,Gh2,Gh3,Gh4,Gh5,Gh6,Gh7, Gh8}. The elementary particle physics as well as its implications for the very early Universe evolution were investigated. The quadratic Weyl action has spontaneous symmetry breaking in a Stueckelberg mechanism, with the result that the Weyl gauge field acquires mass. Hence, one recovers the Einstein-Hilbert action of standard general relativity in the presence of a positive cosmological constant, together with the Proca action for the massive Weyl gauge field \cite{Gh1}. A Weyl-invariant Lagrangian
without ghosts of the form
\bea
L&=& \sqrt{-g}\,\Big\{ - \frac{\xi_j}{2} \, \Big[\frac{1}{6}\, \, \phi^2_j\,R+ g^{\mu\nu} \,
\partial_\mu\phi_j\, \partial_\nu\phi_j\Big]\nonumber\\
&&+(1+\xi_j)\, \frac{1}{2} g^{\mu\nu} \tilde D_\mu\phi_j\,\tilde D_\nu\phi_j
-V(\phi_j)\Big\}.
\eea
was proposed in \cite{Gh2}, where a potential $V(\phi_j)$  for the scalars $\phi_j$ was also added, with $V$ a homogeneous function
$V(\phi_j)=\phi_k^4\, V(\phi_j/\phi_k), k={\rm fixed}$. A successful inflation is possible if one of the scalar fields is identified as the inflaton. Inflation in Weyl gravity coupled to a scalar field leads to results similar to those of the Starobinsky model \cite{Star}, which is recovered for vanishing non-minimal coupling \cite{Gh3}. In \cite{Gh4} it was pointed out that Weyl conformal geometry may play a fundamental role in the early Universe, where the effective theory at short distances becomes conformal. {\it Weyl conformal geometry has a naturally built-in geometric Stueckelberg mechanism, which is broken spontaneously to Riemannian geometry after a particular Weyl gauge transformation} (of gauge fixing). On the other hand,  the Stueckelberg mechanism rearranges the degrees of freedom, conserving their number. Quadratic gravity $R^2+R_{\mu\nu}^2$ in the Palatini formalism, where the connection and the metric are independent, was investigated in \cite{Gh5}. The action has a gauged scale symmetry, or as Weyl gauge symmetry of the Weyl gauge field. In the presence of non-minimally coupled Higgs-like fields, the theory gives successful inflation. A comparative study of inflation in two theories of quadratic gravity with gauged scale symmetry, given by the original Weyl quadratic gravity, and in a theory defined by a similar action but in the Palatini approach obtained by replacing the Weyl connection by its Palatini counterpart, was considered in \cite{Gh6}. In the  absence of matter the Palatini Lagrangian has the form
\bea
L_0=\sqrt{-g} \Big\{ \,\frac{\xi_0}{4!} \, R(\tilde{\Gamma},g)^2-\frac{1}{4\alpha ^2}\,R_{[\mu\nu]}(\tilde{\Gamma})^2\Big\},
\eea
where $\tilde{\Gamma}$ is the Weyl or Palatini connection,
respectively and $\xi_0$ and $\alpha$ are constants. {\it The Einstein-Proca action of the Weyl field, the Planck scale, and the metricity condition emerge in the broken phase, after $\omega _{\mu}$ acquires mass via the Stueckelberg mechanism}. For large Higgs fields inflation is possible. The cosmological evolution of the Weyl conformal geometry and its associated Weyl quadratic gravity was considered in \cite{Gh8}. In the spontaneously broken phase of Weyl gravity Einstein gravity (with a positive cosmological constant) is recovered, after the Weyl gauge field of scale symmetry becomes massive by Stueckelberg mechanism, and decouples. The comparison to the $\Lambda$CDM model shows a very good agreement between these two models for the (dimensionless) Hubble function $h(z)$ and the deceleration parameter $q(z)$ for redshifts $z\leq 3$. Hence, the Weyl conformal geometry and its associated Weyl quadratic gravity may provide an interesting alternative to the $\Lambda$CDM standard model, and to general relativity.

The extremely precise determinations by the Planck satellite of the temperature fluctuations of the Cosmic Microwave Background Radiation \cite{1g,1h}, combined with the observations of the light curves of the distant supernovae \cite{Riess}, have convincingly shown that the Universe is in a phase of accelerating expansion,  of a de Sitter type. Additionally, other important observational results led to the conclusion that in the total matter composition of the Universe baryonic matter represents only around 5\%, while 95\% of matter consists from two mysterious components, called dark matter, and dark energy, respectively. To explain the cosmological observational data, the $\Lambda$CDM model was introduced, which is necessarily based by the inclusion in the Einstein gravitational field equations of the mysterious cosmological constant $\Lambda$, introduced in general relativity in 1917 by Einstein \cite{Ein}, to build-up {\it a static cosmological model}. The $\Lambda$CDM model gives a very good description of the observational data, particularly at low redshifts. But its  theoretical basis is problematic, and there are no satisfactory explanations to the many questions raised by the interpretation and physical nature of $\Lambda$.

Therefore, it is reasonable to assume that in order to obtain a description of the Universe that is both physically and mathematically acceptable, and realistic in its confrontation with observations, {\it one must go outside the limits of standard general relativity}.  Hence, the general relativistic Einstein field equations that give an extremely precise description of the gravitational physics in the Solar System, must be replaced by a new theory of gravity.  One of the possible extensions of general relativity is represented by theories that imply a geometry-matter coupling \cite{Ber,H1, Lm}. For extensive reviews and discussions of theories with geometry-matter coupling see \cite{H2,H3,H4,H5,H6,H7}. Such a geometrical-physical approach leads to gravitational models more complicated than standard general relativity, and they represent an interesting possibility for explaining the accelerating expansion of the Universe, dark energy, and dark matter, respectively. However, these types of theories also raise a number of extremely difficult  physical and mathematical questions.

It is the goal of the present paper to investigate an extension of the conformally invariant Weyl geometric gravity, as introduced in \cite{Gh1,Gh2,Gh3}, by allowing the possibility of a conformally invariant coupling between matter and curvature in a Weyl geometric framework. Under the assumption that the matter Lagrangian is conformally invariant, the simplest possibility of a curvature-matter coupling consists in the addition to the gravitational action of a term of the form $L_m\tilde{R}^2$, where $\tilde{R}$ is the Weyl scalar constructed from the contractions of the Weyl curvatures. This term is conformally invariant, and thus the full conformal invariance of the theory is preserved. After introducing the gravitational action in Weyl geometry, with the help of {\it an auxiliary scalar field}, the action can be linearized in the curvature in Riemann geometry, where it takes the form of a $f\left(R,L_m\right)$ theory, with the matter Lagrangian coupled with the Ricci scalar. We obtain the field equations of the theory by varying the action with respect to the metric tensor. The divergence of the matter energy-momentum tensor turns out to be generally non-zero, indicating that the motion of massive particles is nongeodesic. The evolution equations of the Weyl vector are formulated in analogy with the Maxwell equations of classical electromagnetism, by introducing the electric and magnetic Weyl vectors, thus leading to an electromagnetic type system also containing the matter Lagrangian. The Newtonian limit of the theory is also considered, and the generalized Poisson equation is obtained, which allows to consider the corrections of the Newtonian gravitational potential in the vacuum coming from the Weyl geometry. As a possible astrophysical test of the theory we investigate the motion of the planets in the Newtonian approximation by using an approach based on the use of the Runge-Lenz vector. We interpret the non-conservation of the matter energy-momentum tensor as describing particle production due to the geometry-matter coup[ling, and we obtain the basic thermodynamic parameters (particle number balance, creation pressure, entropy and temperature) of this process by using the thermodynamics of irreversible processes in open systems. The matter creation processes are essentially controlled by the Weyl vector.

In order to consider the cosmological applications of the theory we have performed a {\it spatial averaging of the Weyl vector, and of the Weyl electric and magnetic fields}. As a result of averaging the cosmological effects of Weyl geometry can be described in terms of an effective radiation like fluid, with time dependent only pressure and density satisfying the radiation equation of state. As a result of the averaging of the cosmological field equations one obtains a system of generalized Friedmann equations, describing the evolution of the Universe in a homogeneous and isotropic Friedmann-Lemaitre-Robertson-Walker geometry. The system of cosmological equations is studied numerically, and its solutions are fitted with the observational data, thus allowing to obtain the optimal values of the model parameters.

The present paper is organized as follows. After a brief review of Weyl geometry, the action of the {\it conformally invariant} $f\left(R,L_m\right)$ theory is introduced in Section~\ref{sect1}. After linearizing the action, the gravitational field equations as well as the Weyl field equations are derived. The Newtonian limit of the theory, as well as the generalized Poisson equation are also obtained. The divergence of the matter energy-momentum tensor, the expression of the extra-force, as well as the equations of motion of massive test particles are presented in Section~\ref{sect2}. Solar System constraints on the model parameters are inferred from the study of the perihelion precession of the planet Mercury, performed via the use of the Runge-Lenz vector.  The thermodynamic interpretation of the theory is outlined in Section~\ref{sect3}, where it is pointed out that the present theory may involve particle creation processes whose natural description may be done by using the formalism of the thermodynamics of irreversible processes in open systems. The basic thermodynamic parameters (particle creation rates, creation pressures, entropies and temperatures) are obtained in a general form, and in the Newtonian approximation.  The cosmological applications of the theory are presented in Section~\ref{sect4}, where the averaging procedure of the Weyl field is also presented in detail. The averaged generalized Friedmann equations are solved numerically, and their solutions are fitted to the observational data, thus allowing to obtain the optimal values of the model parameters. We discuss and conclude our results in Section~\ref{sect5}.

{\it In this study we use the Landau-Lifshitz \cite{LaLi} sign conventions, and definitions of the geometric quantities}.

\section{Coupling matter and geometry in conformal Weyl spacetimes}\label{sect1}

In the present Section we first briefly review the fundamentals of Weyl geometry to be subsequently used. Then the action of the theory is presented,  and the gravitational field equations are obtained. The field equations of the Weyl vector, written down in a form similar to the electromagnetic Maxwell equations,  and the Newtonian limit of the theory are also considered.

\subsection{Recap of Weyl geometry}

The Weyl geometry is defined by classes of equivalence $\left( g_{\alpha
\beta },\omega _{\mu }\right) $ of the metric $g_{\alpha \beta }$ and of the
Weyl vector gauge field $\omega _{\mu }$, related by the Weyl gauge
transformations \cite{Gh7},
\begin{eqnarray}
\tilde{g}_{\mu \nu } &=&\Sigma ^{n}g_{\mu \nu }=\left[\tilde{g}_{\mu \nu}\right],\tilde{\omega}_{\mu }=\omega
_{\mu }-\frac{1}{\alpha }\partial _{\mu }\ln \Sigma ,  \notag \\
\sqrt{-\tilde{g}} &=&\Sigma ^{2n}\sqrt{-g},\;\;\;\;\;\;\;\;\;\;\;\tilde{\phi}=\Sigma ^{-n/2}\phi ,
\end{eqnarray}
where by $n$ we have denoted the Weyl charge.
Also we can easily obtain
\begin{eqnarray*}
\left[ \tilde{R}_{\mu \nu }\right] &=&1,\left[ \tilde{\Gamma}_{\nu \rho }^{\mu
}\right] =1,\left[ \tilde{R}\right] =\frac{1}{\Sigma ^n},\left[ \tilde{R}_{\nu \rho \sigma
}^{\mu }\right] =1,\left[ F_{\mu \nu }\right] =1,\nonumber\\
\left[ L_{m}\right] &=&1,\left[ T_{\mu \nu }\right] =\Sigma^n ,\left[ T^{\mu \nu
}\right] =\Sigma ^{-n},\left[ \rho \right] =1,\left[ p\right] =1,\nonumber\\
\left[ T%
\right] &=&1, \left[ u_{\mu }\right] =\Sigma ^{n/2},\left[ u^{\mu }\right] =\Sigma ^{-n/2},%
\left[ j^{\mu }\right] =\Sigma ^{-n/2}.
\end{eqnarray*}

In the above equations, and further, the square brackets $[...]$
denote the degree of $\Sigma$ in the conformal transformation of the physical and geometrical quantities.
Moreover,  $\rho $ denotes the matter energy density, $p$ is
the thermodynamic pressure, $T_{\mu \nu }$ is the ordinary matter
energy-momentum tensor, $T=-\rho +3p$ is the trace of the energy-momentum
tensor, $u_{\mu }$ is the four-velocity of the matter, and $j_{\mu }=\rho
u_{\mu }$ is the matter current, respectively. $L_{m}$ denotes {\it the Lagrangian density of the ordinary (baryonic) matter}, which can be
taken equivalently as $L_{m}=\rho $, or $L_{m}=-p$.

The Weyl gauge vector field is determined by the Weyl connection $\tilde{\Gamma}
$, which can be obtained as a solution of the equations
\begin{equation}\label{5}
\tilde{\nabla}_{\lambda }g_{\mu \nu }=-n\alpha \omega _{\mu }g_{\mu \nu },
\end{equation}%
where $\alpha$ is the Weyl gauge coupling, and
\begin{equation}
\tilde{\nabla}_{\lambda }g_{\mu \nu }=\partial _{\lambda }g_{\mu \nu }-%
\tilde{\Gamma}_{\nu \lambda }^{\rho }g_{\rho \mu }-\tilde{\Gamma}_{\mu
\lambda }^{\rho }g_{\nu \rho }.
\end{equation}

The Weyl geometry is {\it non-metric}, and Eq.~(\ref{5}) can be reformulated in an equivalent form as
\begin{equation}
\left( \tilde{\nabla}_{\lambda }+n\alpha \omega _{\lambda }\right) g_{\mu
\nu }=0.
\end{equation}

Similarly to gauge theory in elementary particle physics, one can construct gauge invariant expressions in which the partial derivative is replaced by a
Weyl covariant derivative, like, for example,  in
\begin{equation}
\partial _{\lambda }\rightarrow \partial _{\lambda }+\mathrm{weight}\times
\alpha \times \omega _{\lambda }.
\end{equation}

Using the permutation of indices and combining the resulting relations, from Eq.~(\ref{5}) we obtain
\begin{equation}
\tilde{\Gamma}_{\mu \nu }^{\lambda }=\Gamma _{\mu \nu }^{\lambda }+\alpha
\frac{n}{2}\left( \delta _{\mu }^{\lambda }\omega _{\nu }+\delta _{\nu
}^{\lambda }\omega _{\mu }-\omega ^{\lambda }g_{\mu \nu }\right) ,
\label{1a}
\end{equation}%
where
\begin{equation}
\Gamma _{\lambda ,\mu \nu }=\frac{1}{2}\left( \partial _{\nu }g_{\lambda \mu
}+\partial _{\mu }g_{\lambda \nu }-\partial _{\lambda }g_{\mu \nu }\right) ,
\end{equation}%
is the Levi-Civita metric connection, and
\begin{equation}
\tilde{\Gamma}_{\mu \nu }^{\lambda }=g^{\lambda \sigma }\tilde{\Gamma}%
_{\lambda ,\mu \nu }.
\end{equation}

Taking the trace in Eq. (\ref{1a}) gives
\begin{equation}
\tilde{\Gamma}_{\mu }=\Gamma _{\mu }+2n\alpha \omega _{\mu }.
\end{equation}

We also introduce the field strength $\tilde{F}_{\mu \nu }$ of $\omega _{\mu
}$, defined according to
\begin{equation}
\tilde{F}_{\mu \nu }=\tilde{\nabla}_{\mu }\omega _{\nu }-\tilde{\nabla}_{\nu
}\omega _{\mu }=\partial _{\mu }\omega _{\nu }-\partial _{\nu }\omega _{\mu
}.
\end{equation}

We can compute the curvatures in Weyl geometry by using the Weyl connection
as follows,
\begin{equation}
\tilde{R}_{\mu \nu \sigma }^{\lambda }=\partial _{\nu }\tilde{\Gamma}_{\mu
\sigma }^{\lambda }-\partial _{\sigma }\tilde{\Gamma}_{\mu \nu }^{\lambda }+%
\tilde{\Gamma}_{\rho \nu }^{\lambda }\tilde{\Gamma}_{\mu \sigma }^{\rho }-%
\tilde{\Gamma}_{\rho \sigma }^{\lambda }\tilde{\Gamma}_{\mu \nu }^{\rho },
\end{equation}%
and
\begin{equation}
\tilde{R}_{\mu \nu }=\tilde{R}_{\mu \lambda \nu }^{\lambda },\tilde{R}%
=g^{\mu \sigma }\tilde{R}_{\mu \sigma },
\end{equation}%
respectively. For the Weyl scalar we obtain
\begin{equation}
\tilde{R}=R-3n\alpha \nabla _{\mu }\omega ^{\mu }-\frac{3}{2}\left( n\alpha
\right) ^{2}\omega _{\mu }\omega ^{\mu }.  \label{R}
\end{equation}

$\tilde{R}$ transforms covariantly, and $\sqrt{-g}\tilde{R}^{2}$ is
invariant. Another important quantity, the Weyl tensor is defined as
\begin{eqnarray}
\tilde{C}_{\mu \nu \rho \sigma } &=&C_{\mu \nu \rho \sigma }-\frac{n\alpha}{%
	4}\Bigg( g_{\mu \rho }\tilde{F}_{\nu \sigma }+g_{\nu \sigma }\tilde{F}_{\mu
	\rho }-g_{\mu \sigma }\tilde{F}_{\nu \rho } \nonumber\\
&&-g_{\nu \rho }\tilde{F}_{\mu
	\sigma }\Bigg)  -\frac{\alpha n}{2}\tilde{F}_{\mu \nu }g_{\rho \sigma },
\end{eqnarray}
where $C_{\mu \nu \rho \sigma }$  is the Weyl tensor of Riemannian geometry, defined in four dimensions as
\begin{eqnarray}
C_{\mu \nu \rho \sigma }&=&R_{\mu \nu \rho \sigma }+\frac{1}{2}\Bigg( R_{\mu
\sigma }g_{\nu \rho }+R_{\nu \rho }g_{\mu \sigma }-R_{\mu \rho }g_{\nu
\sigma }\nonumber\\
&&-R_{\nu \sigma }g_{\mu \rho }\Bigg) +\frac{1}{6}R\left( g_{\mu \rho }g_{\nu \sigma }-g_{\mu \sigma }g_{\nu
\rho }\right) .
\end{eqnarray}

Hence
\begin{equation}
\tilde{C}_{\mu \nu \rho \sigma }^{2}=C_{\mu \nu \rho \sigma }^{2}+\frac{3}{2}%
\left( \alpha n\right) ^{2}\tilde{F}_{\mu \nu }^{2}.
\end{equation}

The tensor $\sqrt{-g}\tilde{C}_{\mu \nu \rho \sigma }^{2}$ as well as all its components
are invariant with respect to the conformal transformations of the metric.

For $C_{\mu \nu \rho \sigma }^{2}$ we find
\begin{equation}
C_{\mu \nu \rho \sigma }^{2}=R_{\mu \nu \rho \sigma }R^{\mu \nu \rho \sigma
}-2R_{\mu \nu }R^{\mu \nu }+\frac{1}{3}R^{2}.
\end{equation}

In the following for the Weyl charge $n$ we will adopt the value $n=1$ only.

\subsection{Conformal coupling of matter and curvature in Weyl geometry}

By using the basic scalars of the Weyl geometry $\left( \tilde{R},\tilde{F}%
_{\mu \nu }^{2},\tilde{C}_{\mu \nu \rho \sigma }^{2}\right) $, the following
action was proposed in \cite{Gh7} to describe the properties of the
gravitational field,
\begin{equation}  \label{S0}
S_{0}=\int \Big[\frac{1}{4!}\frac{1}{\xi ^{2}}\tilde{R}^{2}-\frac{1}{4}\,%
\tilde{F}_{\mu \nu }^{2}-\frac{1}{\eta ^{2}}\tilde{C}_{\mu \nu \rho \sigma
}^{2}\Big]\sqrt{-g}d^{4}x,
\end{equation}%
where two coupling parameters $\xi $ and $\eta $ have also been introduced.
To obtain a realistic approach of the gravitational phenomena the effect of
the matter must also be considered in the action (\ref{S0}) via a
conformally invariant Lagrangian density $\tilde{L}_{m}$.

In the present
study we consider {\it the simplest possibility for the construction of a
conformally invariant matter Lagrangian}, namely, by assuming that $\tilde{L}%
_{m}= L_{m}\tilde{R}^{2}/4!\gamma ^2$, where $L_{m}$ is the ordinary matter
Lagrangian density, with the property $\left[L_m\right]=1$, and $\gamma $ is a coupling constant. Hence in the
presence of matter the conformally invariant action for gravity in Weyl
geometry takes the form
\begin{widetext}
\begin{eqnarray}\label{S1}
S&=&\int \Big[\frac{1}{4!\xi ^2}\tilde{R}^{2}-\frac{1}{4}\,\tilde{F}_{\mu
\nu }^{2}-\frac{1}{\eta ^{2}}\tilde{C}_{\mu \nu \rho \sigma }^{2}+\frac{1}{%
4!\gamma ^{2}}L_{m}\tilde{R}^{2}\Big]\sqrt{-g}d^{4}x  \notag  \label{S} \\
&=&\int \Big[\frac{1}{4!\xi^2}\left( 1+\frac{\xi ^{2}}{\gamma ^{2}}L_m\right)
\tilde{R}^{2}-\frac{1}{4}\,\tilde{F}_{\mu \nu }^{2}-\frac{1}{\eta ^{2}}%
\tilde{C}_{\mu \nu \rho \sigma }^{2}\Big]\sqrt{-g}d^{4}x.
\end{eqnarray}
\end{widetext}

At this moment we introduce an auxiliary scalar field $\phi _{0}$, according to the definition \cite{Gh7},
\begin{equation}
\tilde{R}^{2}\equiv -2\phi _{0}^{2}\tilde{R}-\phi _{0}^{4}.  \label{R2}
\end{equation}
We then substitute $\tilde{R}^{2}\rightarrow-2\phi _{0}^{2}\tilde{R}-\phi _{0}^{4}$ in Eq.~\eqref{S}.
The variation of the action (\ref{S1}) with respect to $\phi _{0}$ leads to
the equation
\begin{equation}
\phi _{0}\left( \tilde{R}+\phi _{0}^{2}\right) =0,
\end{equation}%
which fixes $\phi _{0}^{2}$ as

\begin{equation}
\phi _{0}^{2}=-\tilde{R},
\end{equation}%
and hence through this identification we recover the original form of the
Lagrangian as defined in the initial Weyl geometry. On the other hand, gravity theories containing higher derivative terms of the form $R^n$, are equivalent to standard general relativistic theories with one extra scalar degree of freedom $\phi_0$. These new scalar degree of freedom also appear generally in
in $f(R)$ theories of gravity due to the redefinition of the variables of the model. Considering a general gravitational action of the form $S=\left(1/2\kappa ^2\right)\int{f(R)\sqrt{-g}d^4x}$, one can introduce a scalar field $\phi$ together with an associated potential $V(\phi)$ according to the transformations $\phi \equiv f_R(R)$, and $V(\phi)\equiv R(\phi)f_R(R)-f(R(\phi))$, respectively, which allows to reformulate the gravitational action as $S=\left(1/2\kappa ^2\right)\int{\left[\phi R-V(\phi)\right]\sqrt{-g}d^4x}$. Hence, at the level of action, $R^2\rightarrow \phi R-V(\phi)$, and thus, after performing explicitly the calculations we obtain the relation (\ref{R2}).

Hence, the $R^2$ type gravitational models have the remarkable property of allowing their linearization in the Ricci scalar via the introduction of the scalar degree of freedom.

By substituting Eq. (\ref{R2}) into the action (\ref{S}), with
the use of Eq. (\ref{R}), we obtain
\begin{eqnarray}
\hspace{-0.5cm}S &=&-\int \Bigg\{\frac{1}{2\xi ^{2}}\left( 1+\frac{\xi ^{2}}{%
\gamma ^{2}}L_{m}\right) \Bigg[\frac{\phi _{0}^{2}}{6}R-\frac{\alpha }{2}%
\phi _{0}^{2}\nabla _{\mu }\omega ^{\mu }  \notag \\
\hspace{-0.5cm} &&-\frac{\alpha ^{2}}{4}\phi _{0}^{2}\omega _{\mu }\omega
^{\mu }+\frac{\phi _{0}^{4}}{12}\Bigg]+\frac{1}{4}\,\tilde{F}_{\mu \nu }^{2}+%
\frac{1}{\eta ^{2}}\tilde{C}_{\mu \nu \rho \sigma }^{2}\Bigg\}\sqrt{-g}%
d^{4}x.  \notag \\
\end{eqnarray}

We perform now a conformal transformation of the metric with the conformal
factor $\Sigma =\phi _{0}^{2}/\left\langle \phi _{0}^{2}\right\rangle $,
where $\left\langle \phi _{0}^{2}\right\rangle $ is the (constant) vacuum
expectation value of the field $\phi _{0}$ \cite{Gh7}. Hence the determinant of the
metric tensor transforms as $\sqrt{-g}=\left( \left\langle
\phi _{0}^{2}\right\rangle ^{2}/\phi _{0}^{4}\right) \sqrt{-\hat{g}}$, $%
\tilde{R}=\left( \phi _{0}^{2}/\left\langle \phi _{0}^{2}\right\rangle
\right) \hat{R}$, and $\omega _{\mu }=\hat{\omega}_{\mu }+\left( 2/\alpha
\right) \partial _{\mu }\phi_0/\phi_0 $, respectively. Moreover, {\it we impose the
gauge condition} $\nabla_{\mu }\hat{\omega}^{\mu }=0$.

For the sake of simplicity in the following {\it we will not write explicitly the hats on the conformally rescaled
geometrical quantities}. Hence, the action of the gravitational theory
generated from Weyl geometry in the presence of geometry-matter couplings
takes the form
\begin{align}\label{Sf1}
S &=-\int \Bigg\{\left( 1+\frac{\xi ^{2}}{\gamma ^{2}}L_{m}\right) \Bigg[%
\frac{1}{2}M_{p}^{2}R-\frac{3\alpha ^{2}}{4 }M_p^2\omega _{\mu }\omega ^{\mu
}  \notag \\
&+\frac{3}{2}\xi ^{2}M_{p}^{4}\Bigg] +\frac{1}{4\delta ^{2}}\,\tilde{F}%
_{\mu \nu }^{2}+\frac{1}{\eta ^{2}}C_{\mu \nu \rho \sigma }^{2}\Bigg\}\sqrt{%
	-g}d^{4}x,
\end{align}%
where we have denoted $M_{p}^{2}=\left\langle \phi _{0}^{2}\right\rangle
/6\xi ^{2}$ and $1/\delta ^{2}=1+6\alpha ^{2}/\eta ^{2}$. {\it All physical and
geometrical quantities in the above action are defined in the Riemann space},
but in order to emphasize the geometric origin of $\tilde{F}_{\mu \nu }$, we
will continue to denote it by a tilde. Furthermore, we rescale the matter Lagrangian by introducing a new effective matter variable ${\cal L}_m$, defined as
\be
{\cal L}_m=1+\frac{\xi ^2}{\gamma ^2}L_m.
\ee
Hence, we finally obtain the action of the conformally invariant $f\left(R,L_m\right)$ theory as
\begin{align}\label{Sf}
S &=-\int \Bigg\{{\cal L}_m \Bigg[%
\frac{1}{2}M_{p}^{2}R-\frac{3\alpha ^{2}}{4 }M_p^2\omega _{\mu }\omega ^{\mu
}  \notag \\
&+\frac{3}{2}\xi ^{2}M_{p}^{4}\Bigg] +\frac{1}{4\delta ^{2}}\,\tilde{F}%
_{\mu \nu }^{2}+\frac{1}{\eta ^{2}}C_{\mu \nu \rho \sigma }^{2}\Bigg\}\sqrt{%
	-g}d^{4}x,
\end{align}%

\subsection{Gravitational field equations}

By varying the gravitational action (\ref{Sf}) with respect to the Weyl vector $%
\omega_{\mu}$, it follows that $\omega _{\mu}$ satisfies {\it the generalized system of
Maxwell-Proca type equations},
\begin{equation}
\nabla _{\mu }\tilde{F}^{\mu \nu }+\frac{3}{%
	2}M_p^2\alpha ^{2}\delta^2\left( 1+\frac{\xi ^{2}}{\gamma ^{2}}L_{m}\right) \omega ^{\nu }=0,
\end{equation}
or, equivalently,
\begin{equation}\label{Proca1}
\nabla _{\mu }\tilde{F}^{\mu \nu }+\frac{3}{%
	2}M_p^2\alpha ^{2}\delta^2\mathcal{L}_m \omega ^{\nu }=0.
\end{equation}

Due to its antisymmetry, the Weyl field strength $\tilde{F}^{\mu \nu }$
satisfies automatically, in Riemann geometry,  the equations
\begin{equation}\label{Proca2}
\nabla _{\sigma }\tilde{F}_{\mu \nu }+\nabla _{\mu }\tilde{F}_{\nu \sigma }+\nabla
_{\nu }\tilde{F}_{\sigma \mu }=0.
\end{equation}

We consider now the variation of the action (\ref{Sf}) with respect to the
metric tensor $g^{\mu \nu }$. The variation of the term$-\sqrt{-g}\tilde{F}_{\mu \nu
}^{2}/4\delta ^{2}=-\sqrt{-g}\tilde{F}_{\mu \nu }\tilde{F}^{\mu \nu }/4\delta ^{2}=-\sqrt{-g}%
\tilde{F}_{\mu \nu }\tilde{F}_{\lambda \sigma }g^{\mu \lambda }g^{\nu \sigma
}/4\delta ^{2}$ gives the electromagnetic type energy-momentum tensor
associated to the Weyl field,
\begin{equation}
\tilde{T}_{\mu \nu }^{(\omega)}=\frac{1}{2\delta ^{2}}\left( -\tilde{F}_{\mu
	\lambda }\tilde{F}_{\nu }^{~\lambda }+\frac{1}{4}\tilde{F}_{\lambda \sigma }%
\tilde{F}^{\lambda \sigma }g_{\mu \nu }\right) .
\end{equation}

The variation of the term
\begin{equation}
S_{C^{2}}=-\frac{1}{\eta ^{2}}\int C_{\mu \nu \rho \sigma }^{2}\sqrt{-g}%
d^{4}x,
\end{equation}
gives,
\begin{equation}
\frac{\delta }{\delta g^{\mu \nu }}S_{C^{2}}=-\frac{4}{\eta ^{2}}\int B_{\mu
	\nu }\delta g^{\mu \nu }\sqrt{-g}d^{4}x,
\end{equation}
where $B_{\mu \nu }$, {\it the Bach tensor}, is given by
\begin{equation}
B_{\mu \nu }=\nabla _{\lambda }\nabla _{\sigma }C_{\mu~ \nu }^{~\sigma ~\lambda
}+\frac{1}{2}C_{\mu~ \nu }^{~\lambda ~\sigma }R_{\lambda \sigma }.
\end{equation}

We consider now the variation of the first term in the action (\ref{Sf}).
For the sake of simplicity we denote
\begin{equation}
K=\frac{1}{2}M_{p}^{2}\left( R-\frac{3\alpha ^{2}}{2}\omega _{\mu }\omega
_{\nu }g^{\mu \nu }+3\xi ^{2}M_{p}^{2}\right) .
\end{equation}

Therefore we immediately obtain
\bea\label{vari}
\frac{\delta }{\delta g^{\mu \nu }}\left[ \sqrt{-g}\mathcal{L}_{m}K\right] &=&%
\frac{\delta }{\delta g^{\mu \nu }}\left( \sqrt{-g}\mathcal{L}_{m}\right) K\nonumber\\
&&+\sqrt{-g}\mathcal{L}_{m}\frac{\delta }{\delta g^{\mu \nu }}K.
\eea

For the first term in the above equation we find
\bea
\frac{\delta }{\delta g^{\mu \nu }}\left( \sqrt{-g}\mathcal{L}_{m}\right) K&=&\frac{\partial }{\partial g^{\mu \nu }}\left( \sqrt{-g}\mathcal{L}_{m}\right) K\delta g^{\mu \nu}\nonumber\\
&&=%
\frac{1}{2}\mathcal{T}_{\mu \nu }K\sqrt{-g}\delta g^{\mu \nu },
\eea
where we have introduced {\it the effective energy-momentum tensor} $\mathcal{T}%
_{\mu \nu }$, defined as
\begin{equation}
\mathcal{T}_{\mu \nu }=-\frac{2}{\sqrt{-g}}\frac{\partial \left( \sqrt{-g}%
\mathcal{L}_{m}\right) }{\partial g^{\mu \nu }}.
\end{equation}

The effective energy-momentum tensor $\mathcal{T}_{\mu \nu }$ is related to
the ordinary energy-momentum tensor $T_{\mu \nu }$, defined according to
\begin{equation}
T_{\mu \nu }=-\frac{2}{\sqrt{-g}}\frac{\partial \left( \sqrt{-g}L_{m}\right)
}{\partial g^{\mu \nu }},
\end{equation}%
by the relation
\begin{equation}
\mathcal{T}_{\mu \nu }=g_{\mu \nu }+\frac{\xi ^{2}}{\gamma ^{2}}T_{\mu \nu }.
\end{equation}%
To obtain the above result we have used the identity $\delta \sqrt{-g}%
=-\left( 1/2\right) \sqrt{-g}g_{\mu \nu }\delta g^{\mu \nu }$ \cite{LaLi}. If the matter Lagrangian does not depend on the derivatives of the metric tensor, the effective energy-momentum tensor is given by
\be\label{51}
\mathcal{T}_{\mu \nu }=\mathcal{L}_{m}g_{\mu \nu}-2\frac{\partial \mathcal{L}_{m}}{\partial g^{\mu \nu}}.
\ee

The variation of the Ricci scalar can be obtained as \cite{Rub}
\begin{align}
\delta R=\delta \left( g^{\mu \nu }R_{\mu \nu }\right) =R_{\mu \nu }\delta
g^{\mu \nu }+g^{\mu \nu }\left( \nabla _{\lambda }\delta \Gamma _{\mu \nu
}^{\lambda }-\nabla _{\nu }\delta \Gamma _{\mu \lambda }^{\lambda }\right) ,
\end{align}
where $\nabla _{\lambda }$ is the covariant derivative with respect to the Riemannian Levi-Civita connection $\Gamma $, associated to the metric $g$. Since the variation of the Christoffel symbols is given by
\begin{equation}
\delta \Gamma _{\mu \nu }^{\lambda }=\frac{1}{2}g^{\lambda \alpha }\left(
\nabla _{\mu }\delta g_{\nu \alpha }+\nabla _{\nu }\delta g_{\alpha \mu
}-\nabla _{\alpha }\delta g_{\mu \nu }\right) ,
\end{equation}
we finally obtain for the variation of the Ricci scalar the expression
\begin{equation}
\delta R=R_{\mu \nu }\delta g^{\mu \nu }+\left(g_{\mu \nu }\nabla _{\alpha}\nabla ^{\alpha }  -\nabla _{\mu }\nabla _{\nu }\right)\delta g^{\mu \nu }.
\end{equation}

Hence, for the variation of the second term in Eq. (\ref{vari}) we find
\begin{eqnarray}\label{af}
&&\sqrt{-g}\mathcal{L}_{m}\frac{\delta }{\delta g^{\mu \nu }}K=\sqrt{-g}%
\mathcal{L}_{m}\times   \nonumber \\
&&\Bigg[\frac{1}{2}M_{p}^{2}R_{\mu \nu }\delta g^{\mu \nu }+\frac{1}{2}%
M_{p}^{2}g_{\mu \nu }\nabla _{\mu }\nabla ^{\mu }\delta g^{\mu \nu }
\nonumber \\
&&-\frac{1}{2}M_{p}^{2}\nabla _{\mu }\nabla _{\nu }\delta g^{\mu \nu }-\frac{%
3\alpha ^{2}}{4}M_{p}^{2}\omega _{\mu }\omega _{\nu }\delta g^{\mu \nu }%
\Bigg].
\end{eqnarray}

After partially integrating the second and the third terms
in Eq~(\ref{af}), and discarding the total derivatives, we obtain
\begin{eqnarray}
\hspace{-0.5cm}&&\sqrt{-g}\mathcal{L}_{m}\frac{\delta }{\delta g^{\mu \nu }}K=\frac{1}{%
2}M_{p}^{2}\Bigg[ \mathcal{L}_{m}R_{\mu \nu } \nonumber\\
\hspace{-0.5cm}&&+\left( g_{\mu \nu
}\nabla _{\alpha }\nabla ^{\alpha }-\nabla _{\mu }\nabla _{\nu }\right)
\mathcal{L}_{m}-\frac{3\alpha ^{2}}{2}\mathcal{L}_m\omega _{\mu }\omega _{\nu }%
\Bigg] \sqrt{-g}\delta g^{\mu \nu }.\nonumber\\
\end{eqnarray}

Therefore, by taking into account all the previously partial results, we obtain the gravitational field equations in the conformally symmetric Weyl geometric type model with geometry-matter  coupling in the form
\begin{align}\label{feq}
&M_{p}^{2}\left[ \mathcal{L}_{m}R_{\mu \nu }+\left( g_{\mu \nu
}\nabla _{\alpha }\nabla ^{\alpha }-\nabla _{\mu }\nabla _{\nu }\right)
\mathcal{L}_{m}-\frac{3\alpha ^{2}}{2}\mathcal{L}_m\omega _{\mu }\omega _{\nu }\right]  \nonumber\\
&-\frac{1}{2}M_{p}^{2}\mathcal{T}_{\mu \nu }\left( R-\frac{3\alpha ^{2}}{2}%
\omega _{\alpha }\omega _{\beta }g^{\alpha \beta }+3\xi ^{2}M_{p}^{2}\right)\nonumber\\
& +\frac{8}{%
	\eta ^{2}}B_{\mu \nu }-2\tilde{T}_{\mu \nu }^{(\omega)} =0.
\end{align}

It is interesting to note that ${\cal T}_{\mu \nu}$ {\it can also be interpreted as an effective metric tensor}, depending on the thermodynamic properties of the matter.  For the trace of the effective energy-momentum tensor we obtain ${\cal T}=4+\left(\xi^2/\gamma ^2\right)T$, where $T=T_{\mu}^{\mu}$ is the trace of the ordinary matter energy-momentum tensor.

In the following {\it we assume} that the matter content of the Universe is represented by a {\it perfect fluid} that can be characterized by two thermodynamic quantities only, the energy density $\rho$, and the thermodynamic pressure $p$, respectively. Hence $T_{\mu \nu}$ is given by \cite{LaLi}
\be
T_{\mu \nu}=\left(\rho c^2+p\right)u_{\mu}u_{\nu}-pg_{\mu \nu},
\ee
where $u_{\mu}$ is the four-velocity of the fluid, satisfying the normalization condition $u_{\mu}u^{\mu}=1$. {\it Since in four dimensions the Bach tensor is trace free}, and $\tilde{T}_{\mu}^{(\omega)\mu}=0$, by taking the trace of the field equations (\ref{feq}) we obtain the scalar relation
\begin{eqnarray}\label{trace}
&&\left( \mathcal{L}_{m}R+3\nabla _{\alpha }\nabla ^{\alpha }\mathcal{L}_{m}-%
\frac{3\alpha ^{2}}{2}\mathcal{L}_m\omega ^{2}\right)\nonumber\\
&&-\frac{1}{2}\mathcal{T}\left( R-%
\frac{3\alpha ^{2}}{2}\omega ^{2}+3\xi ^{2}M_{p}^{2}\right) =0,
\end{eqnarray}
where we have denoted $\omega ^{2}=\omega _{\mu }\omega ^{\mu }$. Equivalently, the trace equation can be written as
  \begin{eqnarray}
&&\left( \mathcal{L}_{m}-\frac{1}{2}\mathcal{T}\right) R+3\nabla _{\alpha
}\nabla ^{\alpha }\mathcal{L}_{m}-\frac{3\alpha ^{2}}{2}\left( \mathcal{L}_m-\frac{1}{2}%
\mathcal{T}\right) \omega ^{2}\nonumber\\
&&-\frac{3\xi ^{2}M_{p}^{2}}{2}\mathcal{T}=0.
\end{eqnarray}

By eliminating the term $\nabla _{\alpha }\nabla ^{\alpha }\mathcal{L}_{m}$ between the field equation (\ref{feq}) and the trace equation (\ref{trace}) we obtain an alternative form of the field equation, given by
\begin{align}
&\frac{1}{2}M_{p}^{2}\left[ \mathcal{L}_{m}\left( R_{\mu \nu }-\frac{1}{3}%
g_{\mu \nu }R\right) -\frac{3\alpha ^{2}}{2}\mathcal{L}_m\left( \omega _{\mu }\omega
_{\nu }-\frac{1}{3}g_{\mu \nu }\omega ^{2}\right) \right]  \nonumber\\
&-\frac{1}{4}M_{p}^{2}\left( \mathcal{T}_{\mu \nu }-\frac{1}{3}g_{\mu \nu }%
\mathcal{T}\right) \left( R-\frac{3\alpha ^{2}}{2}\omega ^{2}+3\xi
^{2}M_{p}^{2}\right) \nonumber\\
&-\frac{1}{2}M_{p}^{2}\nabla _{\mu }\nabla _{\nu }%
\mathcal{L}_{m} +\frac{4}{\eta ^{2}}B_{\mu \nu }-	\tilde{T}_{\mu \nu }^{(\omega)}=0.
\end{align}

The field equations can be written by introducing the Einstein tensor as
\begin{eqnarray}
&& R_{\mu \nu }-\frac{1}{2}g_{\mu \nu }R +\frac{8}{\eta
^{2}M_{p}^{2}\mathcal{L}_m}B_{\mu \nu }+\frac{1}{\mathcal{L}_{m}}\hat{\Sigma}_{\mu \nu }%
\mathcal{L}_{m}\nonumber\\
&&+\frac{1}{2}\left( g_{\mu \nu }-\frac{\mathcal{T}_{\mu \nu }}{%
\mathcal{L}_{m}}\right) R
=-\frac{3}{2}\frac{1}{\mathcal{L}_{m}}\left( \frac{\alpha ^{2}}{2}\omega
^{2}-\xi ^{2}M_{p}^{2}\right) \mathcal{T}_{\mu \nu }\nonumber\\
&&+\frac{3\alpha ^{2}}{2}%
\omega _{\mu }\omega _{\nu }+\frac{2}{M_{p}^{2}}\frac{1}{\mathcal{L}_{m}}%
\tilde{T}_{\mu \nu }^{(\omega)},
\end{eqnarray}
where we have denoted $\hat{\Sigma}_{\mu \nu }=g_{\mu \nu }\nabla _{\alpha
}\nabla ^{\alpha }-\nabla _{\mu }\nabla _{\nu }$.

\subsection{The Weyl vector field equations}

In the conformally invariant version of the $f\left(R,L_m\right)$ theory, the Weyl vector satisfies the two field equations (\ref{Proca1}) and (\ref{Proca2}), respectively. In the following we will reformulate these equations in a form similar to the standard Maxwell equations of the electromagnetic theory. Since $\tilde{F}^{\mu \nu}=g^{\alpha \mu}g^{\beta \nu}\tilde{F}_{\alpha \beta}=g^{\alpha \mu}g^{\beta \nu}\left(\nabla _{\alpha}\omega _{\beta}-\nabla _{\beta}\omega _{\alpha}\right)=\nabla ^{\mu}\omega ^{\nu}-\nabla ^{\nu}\omega ^{\mu}$, we obtain
\bea
\nabla _{\mu }\tilde{F}^{\mu \nu }&=&\nabla _{\mu }\nabla ^{\mu }\omega ^{\nu }-\nabla _{\mu  }\nabla ^{\nu }\omega ^{\mu }\nonumber\\
&=&\nabla _{\mu
}\nabla ^{\mu }\omega ^{\nu }+R_{\beta }^{\nu }\omega ^{\beta}-\nabla ^{\nu }\left( \nabla _{\mu }\omega ^{\mu }\right) ,
\eea
where we have used the definition of the Riemann tensor \cite{LaLi},
\be
\left(\nabla _{\mu }\nabla _{\nu }-\nabla _{\nu }\nabla _{\mu }\right) A^{\alpha }=-A^{\beta }R_{\beta \mu \nu
}^{\alpha },
\ee
and of its contraction,
\be
\left( \nabla _{\mu }\nabla _{\nu }-\nabla _{\nu }\nabla _{\mu }\right) A^{\mu }=-A^{\beta }R_{\beta \nu }.
\ee

Hence it follows that the Weyl vector satisfies the generalized wave equation
\be
\Box \omega ^{\nu}+R_{\beta }^{\nu }\omega ^{\beta}-\nabla ^{\nu }\left( \nabla _{\mu }\omega ^{\mu }\right)+\frac{3}{%
	2}M_p^2\alpha ^{2}\delta^2\mathcal{L}_m \omega ^{\nu }=0.
\ee

With the use of the gauge condition for $\omega _{\mu}$, $\nabla _{\mu}\omega ^{\mu}=0$, we obtain
\be
\Box \omega ^{\nu}+R_{\beta }^{\nu }\omega ^{\beta}+\frac{3}{%
	2}M_p^2\alpha ^{2}\delta^2\mathcal{L}_m \omega ^{\nu }=0.
\ee

 We introduce now the definitions
 \be
 \tilde{F}_{0k}=\partial _t\omega _k-\partial _k\omega _0=\tilde{E}_k, k=1,2,3,
 \ee
 and
 \be
 \tilde{F}_{jk}=\partial _j\omega _k-\partial _k\omega _j=-\tilde{B}_{jk}=-\epsilon _{ijk}\tilde{B}^i,j,k=1,2,3,
 \ee
  where $\epsilon_{ijk}$ is the three-dimensional Levi-Civita symbol. Then Eq.~(\ref{Proca2}) gives the analogue Faraday and Gauss laws for the Weyl field,
 \be\label{66}
 \partial _t\tilde{B}^i+\epsilon ^{ijk}\partial _jE_k=0,\partial _j\tilde{B}^j=0.
 \ee

 By introducing the electric and magnetic type Weyl vectors $\vec{\tilde{E}}=\left(\tilde{E}_1, \tilde{E}_2,\tilde{E}_3\right)$ and  $\vec{\tilde{B}}=\left(\tilde{B}_1, \tilde{B}_2,\tilde{B}_3\right)$, Eqs.~(\ref{66}) can be reformulated as
 \be
 \frac{\partial \vec{\tilde{B}}}{\partial t}+\nabla \times \vec{\tilde{E}}=0,\nabla \cdot \vec{\tilde{B}}=0.
 \ee

 In the case of a diagonal metric the inhomogeneous  equation of the Weyl field, Eq.~(\ref{Proca1}),  can be reformulated in terms of the electric and magnetic fields as
 \be
 \partial_k\left(-g^{jj}g^{00}\sqrt{-g}\tilde{E}_j\right)+\frac{3}{%
	2}M_p^2\alpha ^{2}\delta^2\mathcal{L}_m \omega ^{0 }\sqrt{-g}=0,
 \ee
 and
 \be
 \epsilon _{ijk}\partial _j\left(g^{ii}g^{jj}\sqrt{-g}\tilde{B}^k\right)+\frac{3}{%
	2}M_p^2\alpha ^{2}\delta^2\mathcal{L}_m \omega ^{k }\sqrt{-g}=0,
 \ee
respectively.

\subsection{The Newtonian limit, and the generalized Poisson equation}

In the following we will assume that the coupling constant $\eta $ can take  large enough values, and
therefore {\it we will neglect the Bach tensor in the gravitational field
equations}. Then Eqs.~(\ref{feq}) can be written in the form
\begin{eqnarray}\label{feqN}
R_{\nu }^{\mu } &=&-\frac{1}{\mathcal{L}_{m}}\left( \delta _{\nu }^{\mu
}\nabla _{\alpha }\nabla ^{\alpha }-\nabla ^{\mu }\nabla _{\nu }\right)
\mathcal{L}_{m}\nonumber\\
&&+\frac{1}{2\mathcal{L}_{m}}\mathcal{T}_{\nu }^{\mu }\left( R-%
\frac{3\alpha ^{2}}{2}\omega ^{2}+3\xi ^{2}M_{p}^{2}\right) \nonumber\\
&&+\frac{3\alpha ^{2}}{2}\omega ^{\mu }\omega _{\nu }+\frac{2}{M_{p}^{2}%
\mathcal{L}_{m}}\tilde{T}_{\nu }^{(\omega )\mu }.
\end{eqnarray}

We consider now the weak field and low velocity limit of Eqs.~(\ref{feqN}). In our approach we follow \cite{LaLi}.
Since the motion is slow, $\vec{v}^2<<1$, we can neglect all spacelike components in $u^{\mu
}$, and thus we retain only the time component, so that $u^{0}=u_{0}=1$, and
$u^{i }=0$, $i =1,2,3$. We further assume that matter is
pressureless, $p\ll\rho $, and thus the matter energy-momentum tensor $T_{\nu
}^{\mu }=\rho u_{\nu }u^{\mu }$ has only one non-zero component, $%
T_{0}^{0}=\rho $.  In the Newtonian limit, only the $g_{00}$ metric tensor
component, given by $g_{00}=1+2\varphi $, $g^{00}\approx 1$,  where $\varphi $ is the Newtonian
gravitational potential, differs from the Minkowskian values of the metric $g_{ik}=g^{ik}=-1$, $i,k=1,2,3$ \cite{LaLi}.

In the weak field/small velocity limit we obtain $R_{0}^{0}\approx R\approx \Delta \varphi$ \cite{LaLi}. In this
limit we can also neglect in the field equations all derivatives with
respect to the time.  Moreover,  $\nabla _{\alpha }\nabla ^{\alpha }\mathcal{L}%
_{m}=\left( 1/\sqrt{-g}\right) \partial _{\alpha }\left( \sqrt{-g}g^{\alpha
\beta }\partial _{\beta }\mathcal{L}_{m}\right) \approx -\Delta \mathcal{L}%
_{m}$, where in the Newtonian limit $\sqrt{-g%
}\approx 1$, and $\mathcal{L}_{m}$ does not depend on time. Moreover, in the
following we assume that $L_m=\rho $, $\mathcal{L}_{m}=1+\left(\xi^2/\gamma ^2\right)\rho $, and that the time-like component of the Weyl vector field is dominant, so that $\omega ^2\approx \omega ^0\omega _0$. As for the effective energy-momentum tensor of matter, in the Newtonian limit its $\tau _{00}$ component is given by
\be
\tau_{00}\approx \tau _0^0=1+2\phi +\frac{\xi ^2}{\gamma ^2}\rho .
\ee

We also assume that the components of the Weyl vector have a very small variation in space, at least on the scale of the Solar System, and in the vicinity of the Sun. Hence, it follows that the electric $\tilde{E}_k=-\partial _k\omega_0$ and magnetic $-\epsilon _{ijk}\tilde{B}^i=\partial _j\omega _k-\partial _k\omega _j$ Weyl vectors have negligibly small values in the Solar neighborhood, $\tilde{E}_k\approx 0$, and $\tilde{B}_k\approx 0$. Hence, we also neglect the 00 component of the Weyl field energy-momentum tensor, by taking $\tilde{T}_0^{(\omega )0}\propto \left(\vec{\tilde{E}}^2+\vec{\tilde{B}}^2 \right)\approx 0$.

Hence, under these
approximations, from Eq.~(\ref{feqN}) we obtain
\begin{eqnarray}\label{genPo}
&&\left( 1+\frac{\xi ^{2}}{\gamma ^{2}}\rho \right) \Delta \varphi  =\frac{%
3\xi ^{2}}{\gamma ^{2}}\left( \frac{\alpha ^{2}}{2}\omega ^{2}+\xi
^{2}M_{p}^{2}\right) \rho \nonumber\\
&&+6\left( \xi ^{2}M_{p}^{2}-\frac{\alpha ^{2}}{2}%
\omega ^{2}\right) \varphi +
\frac{2\xi ^{2}}{\gamma ^{2}}\Delta \rho +3\left( \frac{\alpha ^{2}}{2}%
\omega ^{2}+\xi ^{2}M_{p}^{2}\right) .\nonumber\\
\end{eqnarray}

\subsubsection{Corrections to the Newtonian potential}

In the limit of the weak geometry-matter coupling the coupling constant $\gamma $ takes very large values, and thus, in the low density, and vacuum,  limit, the generalized Poisson equation, determining the Newtonian potential in the linear representation of the conformal Weyl geometry with nonminimal matter-geometry is given by
\begin{equation}
\Delta \varphi =6\left( \xi ^{2}M_{p}^{2}-\frac{\alpha ^{2}}{2}\omega
^{2}\right) \varphi +3\left( \frac{\alpha ^{2}}{2}\omega ^{2}+\xi
^{2}M_{p}^{2}\right).
\label{Poieq1}
\end{equation}

In the limit of the vanishing Weyl vector, and of the effective cosmological
constant $\xi ^{2}M_{p}^{2}$, we recover the standard vacuum Poisson
equation for the Newtonian potential $\Delta \varphi =0$, with the
spherically symmetric solution given by $\varphi (r)=-C/r$, where $C$ is an
integration constant. In the following we neglect the term proportional to the potential in the generalized Poisson equation, that is, we assume that the condition
\be
\phi <<\frac{\alpha ^2 \omega ^2/2+\xi ^2 M_p^2}{2\left(\xi ^2M_p^2-\alpha \omega ^2/2\right)}.
\ee
By considering for the gravitational potential its Newtonian expression $\phi (r)=\left|GM_{\odot}/r\right|$, the above constraint can be reformulated as
\be\label{condnew}
r>>r_g\left|\frac{\xi ^2M_p^2-\alpha \omega ^2/2}{\alpha ^2\omega ^2/2+\xi^2M_p^2}\right|,
\ee
where $r_g=2GM_{\odot}\approx 3\times 10^{5}$ cm, is the gravitational radius of the Sun. Assuming $\alpha \omega ^2/2>>\xi ^2M_p^2$, the validity of the approximation is restricted to the range $r>>r_g$, corresponding to the standard Newtonian regime.

Hence, in spherical symmetry Eq. (\ref{Poieq1}) becomes
\begin{equation}
\frac{1}{r^{2}}\frac{d}{dr}\left( r^{2}\frac{d\varphi (r)}{dr}\right) =\frac{3
\alpha ^{2}}{2}\omega ^{2}(r)+3\xi ^{2}M_{p}^{2},
\end{equation}%
and it has the general solution
\begin{equation}
\varphi (r)=-\frac{C}{r}+\frac{3\alpha ^{2}}{2}\int^{r}d\varsigma\frac{1}{\varsigma
^{2}}\int^{\varsigma }\theta ^{2}\omega ^{2}\left( \theta \right) d\theta +\frac{\xi
^{2}M_{p}^{2}}{2}r^{2}.
\end{equation}

If in some finite region of the space-time $\omega ^{2}$ can be approximated
by a constant, then the gravitational potential is given by
\begin{equation}\label{potcorr}
\varphi (r)=-\frac{C}{r}+\frac{1}{2}\left(\frac{\alpha ^{2}\omega
^{2}}{2}+\xi ^{2}M_{p}^{2}\right) r^{2}.
\end{equation}

Hence, the presence of the Weyl vector induces important corrections into the gravitational potential, and these modifications could lead to some observational or experimental tests for the confirmation of the presence of Weyl geometry in the Universe.

\section{Energy-momentum balance, equations of motion and Solar System tests}\label{sect2}

We consider now the covariant divergence of the field equations (\ref{feq}). The divergence of the Bach tensor vanishes identically, $\nabla _{\mu}B^{\mu}_{\nu}=0$. For the divergence of the energy-momentum tensor of the Weyl field we obtain
\bea
&&\nabla _{\mu}\tilde{T}_{ \nu }^{(\omega)\mu}=
\frac{1}{\delta ^{2}}\times \nonumber\\
&&\left( \frac{1}{2}\tilde{F}^{\lambda \sigma }\nabla_{\nu }\tilde{F%
}_{\lambda \sigma }-\tilde{F}^{\mu \lambda }\nabla _{\mu }\tilde{F}%
_{\nu \lambda }-\tilde{F}_{\nu \lambda }\nabla _{\mu }\tilde{F}%
^{\mu \lambda }\right) .
\eea

With the use of Eqs.(\ref{Proca1}) and (\ref{Proca2}), we obtain immediately
\be
\nabla _{\mu}\tilde{T}_{ \nu }^{(\omega )\mu}=\frac32\alpha^2\delta^2 M_p^2\mathcal{L}_m \omega_\nu.
\ee

We proceed now to the calculation of the divergence of the matter energy-momentum tensor. In order to do so, we will use the geometric identity \cite{Koi}
\be
\left(\nabla _{\nu}\Box-\Box \nabla _{\nu}\right)\mathcal{L}_m=-R_{\mu \nu}\nabla ^{\mu}\mathcal{L}_m,
\ee
where $\Box=\nabla _{\alpha}\nabla ^{\alpha}$. By taking now the divergence of the field equation (\ref{feq}), we obtain
\begin{eqnarray}\label{econs}
\nabla _{\mu }\mathcal{T}_{\nu }^{\mu }&=&\left(\mathcal{L}_{m}\delta _{\nu
}^{\mu }-\mathcal{T}_{\nu }^{\mu }\right) \nabla _{\mu }\ln \left( R-\frac{%
3\alpha ^{2}}{2}\omega ^{2}+3\xi ^{2}M_{p}^{2}\right)\nonumber\\
&& -\frac{6\alpha
^{2}\omega_\nu\omega^\mu\nabla_\mu\mathcal{L}_m}{2R-%
3\alpha ^{2}\omega ^{2}+6\xi ^{2}M_{p}^{2}}:= Q_{\nu }.
\end{eqnarray}

Hence, in conformal $f\left(R,L_m\right)$ type theory with geometry-matter coupling generally the matter energy-momentum tensor is not conserved, and the degree of non-conservation is described by the vector $Q_{\nu}$. If under some special conditions $Q_{\nu}=0$, the theory becomes a conservative one.

By taking into account Eq. (\ref{51}), the divergence of the matter
energy-momentum tensor can be written in the equivalent form
\bea\label{bal}
\nabla _{\mu }\mathcal{T}_{\nu }^{\mu }&=&2g^{\mu \alpha }\frac{\partial
\mathcal{L}_{m}}{\partial g^{\alpha \nu }}\nabla _{\mu }\ln \left( R-\frac{%
3\alpha ^{2}}{2}\omega ^{2}+3\xi ^{2}M_{p}^{2}\right) \nonumber\\
&& -\frac{6\alpha
	^{2}\omega_\nu\omega^\mu\nabla_\mu\mathcal{L}_m}{2R-%
	3\alpha ^{2}\omega ^{2}+6\xi ^{2}M_{p}^{2}}.
\eea

\subsection{Energy and momentum balance equations}

To obtain the energy balance equation in conformal $f\left(R,L_m\right)$ gravity we multiply both sides of
Eq.~(\ref{econs}) by $u^{\mu }$. For the left hand side we obtain
\begin{eqnarray}
u_{\mu }\nabla _{\nu }T^{\mu \nu } &=&u_{\mu }\nabla _{\nu }\left(
\rho +p\right) u^{\mu }u^{\nu }+u_{\mu }\left( \rho +p\right) \times
\nonumber \\
&&\nabla _{\nu }(u^{\mu }u^{\nu })-u_{\mu }\nabla ^{\mu }p  \nonumber \\
&=&u^{\nu }\nabla _{\nu }(\rho +p)+\left( \rho +p\right) \times   \nonumber
\\
&&(u_{\mu }u^{\nu }\nabla _{\nu }u^{\mu }+u_{\mu }u^{\mu }\nabla _{\nu
}u^{\nu })-\dot{p}  \nonumber \\
&=&\dot{\rho}+\left( \rho +p\right) \nabla _{\nu }u^{\nu },
\end{eqnarray}%
where we have used the definition $\dot{\rho}=u^{\mu }\nabla _{\mu }\rho =%
\mathrm{d}\rho /\mathrm{d}s$, and the relations $u_{\mu }u^{\mu }=1$, and $%
u_{\mu }u^{\nu }\nabla _{\nu }u^{\mu }=0$, respectively. Hence we have
obtained the energy balance equation for the conformally invariant $f\left(R,L_m\right)$ theory as
\begin{equation}\label{dotrho}
\dot{\rho}+(\rho +p)\nabla _{\mu }u^{\mu }=\frac{\gamma^2}{\xi^2}u_{\mu}Q^{\mu}.
\end{equation}

In the following we denote $\nabla _{\mu }u^{\mu }=3H$.  Hence the energy-balance equation takes the form
\be\label{dotrho1}
\dot{\rho}+3(\rho +p)H=\frac{\gamma^2}{\xi^2}u_{\mu}Q^{\mu}.
\ee

We introduce now the projection operator $h_{\thickspace \lambda}^\nu$, defined according to
\be
 h_{\thickspace \lambda}^\nu \equiv
\delta_{\thickspace \lambda}^\nu - u^\nu u_\lambda,
\ee
with the basic properties
\be
u_\nu h_{\thickspace \lambda}^\nu = 0, h_{\thickspace %
\lambda}^\nu \nabla_\mu u_\nu = \nabla_\mu u_\lambda,
\ee
and
\be
h^{\nu \lambda}
\nabla_\nu = \left( g^{\nu\lambda} - u^\nu u^\lambda \right) \nabla_\nu =
\nabla^\lambda - u^\lambda u^\nu \nabla_\nu,
\ee
respectively. After multiplying Eq.~(\ref{econs}) with $h_{\thickspace \lambda}^\nu$,  we obtain the momentum balance equation of the theory, representing the (non-geodesic) equation of motion of massive test particles, as
\begin{eqnarray}\label{force0}
u^{\nu} \nabla_{\nu} u^{\lambda} &=&\frac{\mathrm{d}^2x^{\lambda}}{\mathrm{d}%
	s^2}+\Gamma_{\mu \nu}^{\lambda }u^{\mu }u^{\nu} = \frac{h^{\nu \lambda} }{\rho + p}\left[\frac{\gamma ^2}{\xi ^2}Q_{\nu}-
\nabla_{\nu} p \right] \nonumber\\
&&:= f^{\lambda},
\end{eqnarray}
where $f^{\lambda}$ can be interpreted as the {\it extra force} acting on massive test particles moving in the Weyl geometry. The extra force satisfies the condition of being perpendicular to the four-velocity, $f^{\lambda}u_{\lambda}=0$. Eq.~(\ref{force0}) indicates that in the presence of the conformally invariant geometry-matter coupling the motion is {\it non-geodesic}, and the dynamical motion of particles become more complicated than in standard general relativity.

\subsection{Variational principle for the momentum balance equation, and the Newtonian limit}

The equation of motion of massive test particle in conformal Weyl gravity with geometry-matter coupling,   Eq.~(\ref{force0}), can be also derived with  the use of the  variational principle
\begin{equation}\label{actpart}
\delta S_{p}=\delta \int L_{p}ds=\delta \int \sqrt{\Phi}\sqrt{g_{\mu \nu
}u^{\mu }u^{\nu }}ds=0,
\end{equation}
where $S_{p}$ and $L_{p}=\sqrt{\Phi}\sqrt{g_{\mu \nu }u^{\mu }u^{\nu }}$ are
the action and the Lagrangian density for test particles, respectively,
with $\sqrt{\Phi}$ a yet unknown quantity that must be obtained through comparison with the equation of motion (\ref{force0}).

To prove this result we begin with the Euler-Lagrange equations that give the trajectories of
the action~(\ref{actpart}),
\begin{equation}
\frac{d}{ds}\left( \frac{\partial L_{p}}{\partial u^{\lambda }}\right) -%
\frac{\partial L_{p}}{\partial x^{\lambda }}=0.
\end{equation}

We obtain successively
\begin{equation}
\frac{\partial L_{p}}{\partial u^{\lambda }}=\sqrt{\Phi}u_{\lambda }
\end{equation}
and
\begin{equation}
 \frac{\partial L_{p}}{\partial x^{\lambda }}=\frac{1}{2} \sqrt{\Phi}g_{\mu \nu,\lambda }u^{\mu }u^{\nu }+\frac{ 1}{2} \frac{\Phi_{,\lambda }}{\sqrt{\Phi}},
\end{equation}
respectively. Then after a simple calculation we find the equations of motion of the test particles as given by
\begin{equation}\label{mota}
\frac{d^{2}x^{\lambda }}{ds^{2}}+\Gamma _{\mu \nu }^{\lambda }u^{\mu }u^{\nu
}+\left( u^{\lambda }u^{\nu }-g^{\lambda \nu }\right) \nabla _{\nu }\ln \sqrt{\Phi}=0.
\end{equation}

By comparing Eq.~(\ref{mota})  with the equation of motion Eq.~(\ref{force0}), it turns out
that the explicit form of $\sqrt{\Phi}$ can be obtained as a solution of the equation,
\be\label{Phi}
\nabla _{\nu }\ln \sqrt{\Phi}=  \frac{1 }{\rho + p}\left[\frac{\gamma ^2}{\xi ^2}Q_{\nu}-
\nabla_{\nu} p \right].
\ee
 When $Q_{\nu}\rightarrow 0$, we recover the geodesic equation of the standard general relativistic motion. If we adopt for $L_m$ the expression $L_m=\rho$, then
\be
\delta L_m=\delta \rho=\frac{1}{2}\left(\rho +p\right)h_{\mu \nu}\delta g^{\mu \nu},
\ee
and
\be
\frac{\partial \mathcal{L}_m}{\partial g^{\mu \nu}}=\frac{\xi ^2}{2\gamma ^2}\left(\rho +p\right)h_{\mu\nu},
\ee
respectively, and the energy-momentum balance equation (\ref{bal}) takes the form
\bea
\nabla _{\mu }\mathcal{T}_{\nu }^{\mu }&=&\frac{\xi ^2}{\gamma ^2}\left(\rho+p\right)h^{\mu}_{\nu}\nabla _{\mu }\ln \left( R-\frac{%
3\alpha ^{2}}{2}\omega ^{2}+3\xi ^{2}M_{p}^{2}\right) \nonumber\\
&& -\frac{6\alpha
	^{2}\omega_\nu\omega^\mu\nabla_\mu\mathcal{L}_m}{2R-%
	3\alpha ^{2}\omega ^{2}+6\xi ^{2}M_{p}^{2}}.
\eea

In the static spherically symmetric limit, and in the case of a pressureless fluid, we obtain
\be
\frac{\gamma ^2}{\xi ^2}Q_{\nu}\approx \rho h^{\mu}_{\nu}\nabla _{\mu }\ln \left( R-\frac{%
3\alpha ^{2}}{2}\omega ^{2}+3\xi ^{2}M_{p}^{2}\right).
\ee
 Under the assumption of small velocities and weak gravitational fields, the only surviving component of the projection operator is $h_1^1=1$, and thus from Eq.~(\ref{Phi}) we obtain the function $\sqrt{\Phi}$ as given by
 \be
 \sqrt{\Phi}\approx  R-\frac{%
3\alpha ^{2}}{2}\omega ^{2}+3\xi ^{2}M_{p}^{2}.
 \ee

 \subsubsection{The Newtonian limit of the equation of motion}

The variational principle given by Eq~(\ref{actpart}) can also be used to investigate the Newtonian
limit of the equations of motion of the test particles in the conformally invariant Weyl gravity with geometry-matter coupling. In the limit of the weak gravitational fields and low particle velocities, the space-time metric can be approximated as
\begin{equation}
ds\approx \sqrt{1+2\phi -\vec{v}^{2}}dt\approx \left( 1+\varphi -\frac{\vec{v}
^{2}}{2}\right) dt,
\end{equation}
where $\varphi $ is the Newtonian potential, and $\vec{v}$ is
the usual tridimensional velocity of the fluid.

Taking into account the first order of approximation, the equations of motion of the particle in a weak gravitational field can be derived as
\begin{equation}
\vec{a}=\frac{d\vec{v}}{dt}=-\nabla \varphi=\vec{a}_{N}+\vec{a}_{E},
\end{equation}
where $\vec{a}$ is the total acceleration of the system and $\vec{a}%
_{N}=-\nabla \left(C/r\right)=C/r^2$ is the Newtonian gravitational acceleration, while $\vec{a}_E$ is the extra-acceleration induced by the presence of Weyl geometric effects. By taking into account Eq.~(\ref{potcorr}) we obtain for the extra-acceleration induced due the coupling between matter and curvature in Weyl gravity the expression
 \be\label{103}
 \left|\vec{a}_E\right|\approx \left(\frac{\alpha ^2}{2}\omega ^2+\xi ^2M_p^2\right)r.
 \ee

 It is interesting to note that the extra-acceleration induced by the conformal gravity model of Weyl geometry has the opposite sign with respect to the Newtonian acceleration, that is, it has a repulsive effect.

\subsection{Solar System tests of conformal $f\left(R,L_m\right)$ gravity}

To obtain an estimate of the effects of the extra-force in conformal $f\left(R,L_m\right)$ gravity, induced by the coupling
between geometry and matter, we investigate the orbital parameters of the motion of the
planets around a central massive object (the Sun). The motion of massive test particles in a gravitational field can be studied in a simple way with the help of the Runge-Lenz vector, defined as \cite{prec,prec1}
\be
\vec{A}=\vec{v}\times \vec{L}-\alpha_0 \vec{e}_{r},
\ee
where $\vec{v}$ is the velocity of the planet  of mass $m$ with respect to the Sun,  $\mu =mM_{\odot }/\left( m+M_{\odot}\right) $ is the reduced mass, $\alpha_0 =GmM_{\odot}$, $\vec{p}=\mu \vec{v}$ is the relative
momentum, $\vec{r}=r\vec{e}_{r}$ is the two-body position vector,  and
\be
\vec{L}=\vec{r}\times \vec{p}=\mu r^{2}\dot{\theta}\vec{k},
\ee
 is the angular momentum of the planet, respectively. In the following $M_{\odot}$ denotes the Solar mass.

Alternatively the Runge-Lenz vector can be represented  as,
\be
\vec{A}=\left( \frac{\vec{L}^{2}}{\mu r}-\alpha_0 \right) \vec{e}_{r}-%
\dot{r}L\vec{e}_{\theta }.
\ee
The derivative of $\vec{A}$  with respect to the polar
angle $\theta $ can be obtained as \cite{prec,prec1},
\be
\frac{d\vec{A}}{d\theta} =r^{2}\left[ \frac{dV(r)}{dr}-\frac{\alpha_0}
{r^{2}}\right] \vec{e}_{\theta },
\ee
where $V(r)$ is the potential of the central force.

 In the case of an elliptical orbit of
eccentricity $e$, period $T$, and major semi-axis $a$, the equation of the trajectory of the test particle is given in the Newtonian approximation by \cite{LaLim}
\be
\left( \frac{L^{2}}{\mu \alpha_0} \right) r^{-1}=1+e\cos \theta .
\ee
Since at the perihelion $\dot{r}=0$ and $\theta =0$, for the magnitude of the Runge-Lenz vector we obtain $\vec{A}=e\alpha _0\vec{i}$, and $\left|\vec{A}\right|^2=e^2\alpha _0^2$, respectively.

We model the gravitational field in the Solar System by a potential term consisting  of two components, the Post-Newtonian potential,
\be
V_{PN}(r)=-\frac{\alpha_0} {r}-3\frac{\alpha_0 ^{2}}{mr^{2}},
\ee
and the extra contribution arising from the conformal coupling between geometry and matter.
Hence we obtain first
\be
\frac{d\vec{A}}{d\theta} =r^{2}\left[ 6\frac{\alpha_0 ^{2}}{mr^{3}}+m\vec{a}%
_{E}(r)\right] \vec{e}_{\theta },
\ee
where we have assumed that $\mu \approx m$, an approximation that works very well within the range of the Solar System. For a variation of $\theta $ of $2\pi $, the change in the direction $\Delta \tilde{\phi} $ of
the perihelion of a planet is given by
\be
\Delta \tilde{\phi}\vec{k} =\frac{ \vec{A}\times \frac{d%
\vec{A}}{d\theta}}{\vec{A}^2},
\ee
and it is found in a general form as
\be
\Delta \tilde{\phi} = \frac{1}{\alpha_0 e} \int_{0}^{2\pi }\left\vert \vec{A}\times \frac{d%
\vec{A}}{d\theta} \right\vert d\theta .
\ee
Explicitly, the perihelion precession is obtained as,
\begin{align}\label{prec}
\Delta \tilde{\phi} &=24\pi ^{3}\left( \frac{a}{T}\right) ^{2}\frac{1}{1-e^{2}}+\frac{%
L}{8\pi ^{3}me}\frac{\left( 1-e^{2}\right) ^{3/2}}{\left( a/T\right) ^{3}}\times \nonumber\\
&\int_{0}^{2\pi }\frac{a_{E}\left[ L^{2}\left( 1+e\cos \theta \right)
^{-1}/m\alpha_0 \right] }{\left( 1+e\cos \theta \right) ^{2}}\cos \theta
d\theta ,
\end{align}
where we have used the relation $\alpha_0 /L=2\pi \left( a/T\right) /\sqrt{%
1-e^{2}}$ \cite{LaLim}. The first term of the above relation gives the expression of the standard general relativistic precession of the perihelion of the planets, The second term represents the contribution to the perihelion precession of the planets due to the presence of the extra-force generated by the Weyl geometric effects.

We consider now the application of Eq.~(\ref{prec}) to the simple the case in which the Weyl type geometric acceleration can be considered (approximately) as a constant,  $\vec{a}_E\approx$ constant, an approximation that could be valid for small regions of the space-time.
 By assuming that  $\vec{a}_E$ is a constant, from of Eq.~(\ref{prec}) we obtain for the perihelion precession of a planet
the expression
\begin{equation}\label{prec1}
\Delta \tilde{\phi} =\frac{6\pi GM_{\odot}}{a\left( 1-e^{2}\right) }+\frac{2\pi a^{2}%
\sqrt{1-e^{2}}}{GM_{\odot}}a_{E},
\end{equation}
where Kepler's third law, $T^2=4\pi ^2a^3/GM_{\odot}$ was also used.

We apply now Eq.~(\ref{prec1}) to the case of the planet Mercury, for which $a=57.91\times 10^{11}$ cm, and $e=0.205615$,
respectively. For the mass of the Sun we adopt the value $M_{\odot }=1.989\times 10^{33}$ g.  With the use of these
numerical values from the first term in Eq. (\ref{prec1}) we obtain the standard
general relativistic value for the precession angle of Mercury, $\left( \Delta \tilde{\phi}
\right) _{GR}=42.962$ arcsec per century.  On the other hand the observed value of the planet's precession is $\left(\Delta \tilde{\phi} \right)_{obs}=43.11\pm0.21$ arcsec per century \cite{merc, merc1}.

Hence, the difference $\left(\Delta \tilde{\phi} \right)_{E}=\left(\Delta \tilde{\phi} \right)_{obs}-\left( \Delta \tilde{\phi}
\right) _{GR}=0.17$ arcsec per century can be interpreted as coming from other physical effects. Therefore, the observational constraints imply
that the value of the constant extra-acceleration $a_E$ must satisfy the condition $a_E\leq 1.28\times 10^{-9}$ cm/s$^2$. This value of $a_E$, {\it obtained from the
Solar System observations under the assumption of its constancy}, is somewhat smaller than the value of the extra-acceleration $%
a_{0}\approx 10^{-8}$ cm/s$^{2}$, necessary to explain the "dark matter" properties. By using the expression of the Weyl type extra-acceleration as given by Eq.~(\ref{103}), we obtain the following constraint on the Weyl gravity parameters $\left(\alpha, \xi, \omega\right)$,
\be
\left.\left(\frac{\alpha }{2}\omega ^2+\xi ^2M_p^2\right)\right|_{{\rm Mercury}}\leq 2.455\times{10}^{-43}\;{\rm cm}^{-2}.
\ee
It is interesting to note that the term $\xi ^2M_p^2$, corresponding to the cosmological constant, gives in the above relation a contribution of the order of $10^{-56}\;{\rm cm}^{-2}$, and thus, to fully explain the Mercury perihelion precession, one cannot rule out the possibility of the presence of some extra gravitational effects acting at both the Solar System and galactic levels. Of course, the assumption of a constant extra-force may not be correct on larger astronomical scales, and thus a full confrontation of Weyl geometric gravity with observations may require a more complicated approach.

\section{Thermodynamic interpretation of the conformal $f\left(R,L_m\right)$ gravity theories}\label{sect3}

For the sake of completeness we briefly discuss a possible thermodynamic
interpretation of the Weyl geometry inspired conformal quadratic $f\left(R,L_m\right)$ gravity theory.
The non-conservation of the matter energy-momentum tensor in the Weyl geometric background strongly suggests
that, due to the specific conformal matter-curvature coupling in the present version of the $f\left(R,L_m\right)$ theory, particle generation processes may
take place on a microscopic scale, both locally, as well as during the cosmological evolution. It is interesting to note that this effect is also
present in quantum field theories in curved space-times, as discussed extensively in
\cite{Parker,Parker1,Star1, Parker2}. In quantum field theory particle creation from the gravitational field is a consequence of the time variation
of the field. In particular, the finite regularized average value of the energy-momentum tensor of a quantum scalar field in anisotropic cosmology, including both particle creation and vacuum polarization, was obtained in \cite{Star1}. However, in conformal $f\left(R,L_m\right)$ gravity this phenomenon is general, and not restricted to time variability.  Hence, conformal $f\left(R,L_m\right)$ theory, which naturally contain matter
creation, could lead to an effective semiclassical equivalent description of
quantum field theoretical processes in gravitational fields. In the following in the discussion of particle creation we use the approach introduced in \cite{Ha14}.

\subsubsection{Thermodynamic quantities in the presence of particle creation}

The presence of particles creation in a classical physical theory is a direct consequence of the fact that the covariant divergences of
the basic thermodynamic and physical quantities, including the energy-momentum tensor, the  particle, energy and entropy fluxes, respectively,
do not vanish. In this case, the balance equilibrium equations must be reformulated in order to
explicitly include particle creation processes in the basic evolution equations \cite{P-M,Lima,Su}. The balance equation for the particle flux
$N^{\mu} \equiv nu^{\mu}$, where $n$ is the particle number density, becomes in the presence of particle creation from the gravitational field,
\begin{equation}
\nabla _{\mu}N^{\mu}=\dot{n}+3Hn=n\Theta,
\end{equation}
where $\Theta $ is the particle creation rate. If the condition $\Theta \ll H$ is satisfied, matter creation can be neglected in any physical model. We also introduce the entropy flux vector $S^{\mu}$, defined as $S^{\mu} \equiv \tilde{s}u^{\mu} = n\sigma u^{\mu}$, where by $\tilde{s}$ we denote the entropy density, while $\sigma $ represents the entropy per particle. In the presence of particle creation weIf  obtain the divergence of the entropy flux as
\begin{equation}\label{62b}
\nabla _{\mu}S^{\mu}=n\dot{\sigma}+n\sigma \Theta\geq 0.
\end{equation}
If $\Theta =0$, the entropy is conserved, and the thermodynamic process is adiabatic. For a particular thermodynamic model in which $\sigma $ can be taken as constant, we find
\begin{equation}\label{con1}
\nabla _{\mu}S^{\mu}=n\sigma \Theta =\tilde{s}\Theta \geq 0.
\end{equation}
The above relation indicates  that if the entropy per particle is a constant, the variation of the entropy is exclusively determined by the  gravitational matter production effects. By taking into account that $\tilde{s}>0$, from Eq.~(\ref{con1}) it follows that the particle production rate $\Theta$ must satisfy the condition $\Theta \geq 0$. Hence, particles can be generated from gravitational fields,  but the opposite process is forbidden. In the presence of matter production from gravity the energy-momentum tensor of a fluid  must be also adjusted to make it consistent with the second law of thermodynamics. This can be done by adding to the equilibrium component $T^{\mu \nu}_\text{eq}$ a new term $\Delta T^{\mu \nu}$, so that \cite{Bar}
\begin{equation}\label{64}
T^{(tot)\mu \nu}=T^{\mu \nu}_\text{eq}+\Delta T^{\mu \nu},
\end{equation}
where $\Delta T^{\mu \nu}$ denotes the adjustment due to
particle production. In an isotropic and homogeneous geometry, $\Delta T^{\mu \nu}$ should be representable
via a single scalar quantity. Hence, generally we can write \cite{Bar}
\begin{equation}
\Delta T_{\; 0}^0=0, \quad \Delta T_{\; i}^j=-P_c\delta_{\; i}^j,i,j=1,2,3,
\end{equation}
where $P_c$ denotes the \textit{creation pressure}, a dynamical quantity that describes
phenomenologically the effects of the gravitational matter production on a
macroscopic thermodynamic system. In a fully covariant representation we have \cite{Bar}
\begin{equation}
\Delta T^{\mu \nu}=-P_ch^{\mu \nu}=-P_c\left(g^{\mu \nu}-u^{\mu}u^{\nu}\right),
\end{equation}
a relation from which we immediately obtain $u_{\mu}\nabla _{\nu}\Delta T^{(tot)\mu \nu}=3HP_c$.
Hence, in the presence of matter production, the total energy balance
equation $u_{\mu}\nabla _{\nu}T^{(tot)\mu \nu}=0$, which follows from Eq.~\eqref{64}, immediately gives
\begin{equation}
\dot{\rho}+3H\left(\rho+P+P_c\right)=0.
\end{equation}

The thermodynamic quantities must also satisfy the Gibbs law, which can be
formulated as \cite{Lima}
\begin{equation}
n \tilde{T} \mathrm{d} \left(\frac{\tilde{s}}{n}\right)=n\tilde{T}\mathrm{d}\sigma=\mathrm{d}\rho -\frac{\rho+p}{n}\mathrm{d}n,
\end{equation}
where $\tilde{T}$ is the thermodynamic temperature of the system.

\subsubsection{Conformal $f\left(R,L_m\right)$ gravity and irreversible thermodynamics}

With the use of some simple algebraic transformations the energy balance equation~\eqref{dotrho1} of the conformally invariant $f\left(R,L_m\right)$ theory can be rewritten as
\begin{equation}\label{76}
\dot{\rho}+3H\left( \rho +P+P_{c}\right) =0,
\end{equation}%
where the creation pressure $P_{c}$ is defined as
\begin{eqnarray}
P_c=-\frac{\gamma^2}{3\xi^2H}u_{\mu}Q^{\mu}.
\end{eqnarray}
 Then the generalized energy balance Eq.~\eqref{76}
can be obtained by taking the divergence of the total energy momentum tensor %
$T^{\mu \nu }$ of the conformal $f\left(R,L_m\right)$ gravity, defined as
\begin{equation}
T^{\mu \nu }=\left( \rho +P+P_{c}\right) u^{\mu }u^{\nu }-\left(
P+P_{c}\right) g^{\mu \nu }.
\end{equation}

By assuming that particle creation is an adiabatic process,
with $\dot{\sigma}=0$, the Gibbs law gives
\begin{equation}
\dot{\rho}
=\left(\rho+p\right)\frac{\dot{n}}{n}
=\left(\rho+P\right)\left(\Theta-3H\right).
\end{equation}

With the use of the energy balance equation we obtain immediately the
relation between the matter production rate and the creation pressure as
\begin{equation}
\Theta=\frac{-3HP_c}{\rho+p}=\frac{\gamma^2}{\xi^2}\frac{u_{\mu}Q^{\mu}}{\rho +p}.
\end{equation}
Hence in the conformal  $f\left(R,L_m\right)$ gravity theory for the particle
creation rate we find the general expression
\begin{eqnarray}
\Theta =-\frac{6\alpha^2\omega_\mu u^\mu\omega_\nu\nabla^\nu L_m}{(\rho+p)(2R-3\alpha^2\omega^2+6\xi^2M_p^2)}.
\end{eqnarray}
In obtaining the above expression, we have used the relation $u_{\mu}h^{\mu \nu}=0$, and
\begin{align}
\mathcal{L}_m\delta^\mu_\nu-\mathcal{T}^\mu_\nu=(\rho+p)h^\mu_\nu,
\end{align}
which is true for a perfect fluid with Lagrangian $L_m=\rho$.
The condition $\Theta \geq 0$ imposes an important constraint on the
physical parameters of the theory. By assuming for the matter content of the theory pressureless dust,
with $p=0$, we obtain the following
general cosmological constraint that must be satisfied
for all times in the linear/scalar representation of the conformal $f\left(R,L_m\right)$ theory,
\begin{equation}
\frac{u_{\mu}Q^{\mu}}{\rho +p}\geq 0.
\end{equation}

The divergence of the entropy flux vector can be obtained in terms of
the creation pressure as
\begin{equation}
\nabla _{\mu}S^{\mu}=\frac{-3 n \sigma H P_c}{\rho +p}=\frac{\gamma^2}{\xi^2}\frac{n\sigma }{\rho+p}u_{\mu}Q^{\mu}.
\end{equation}

\subsubsection{The temperature evolution}

We consider now the evolution of the temperature in an open thermodynamic system with matter
production. To consistently describe the time evolution of the relativistic
fluid, we introduce the two parametric equations of state for the matter density and pressure, which
generally are given by $\rho =\rho (n, \tilde{T} )$ and $p=p(n,\tilde{T})$, respectively. Then
we immediately find
\begin{equation}
\dot{\rho}=\left(\frac{\partial \rho }{\partial n} \right)_{\tilde{T}}\dot{n}+\left(%
\frac{\partial \rho }{\partial \tilde{T}} \right)_n\dot{\tilde{T}}.
\end{equation}

With the use of the energy and particle balance equations we obtain
\bea\label{78a}
-3H\left(\rho +p+P_c\right)&=&\left(\frac{\partial \rho }{\partial n}
\right)_{\tilde{T}} n\left(\Theta-3H\right)\nonumber\\
&&+\left(\frac{\partial \rho }{\partial \tilde{T}} \right)_n\dot{\tilde{T}}.
\eea

With the use of the thermodynamic identity \cite{Bar}
\begin{equation}\label{Termid}
\tilde{T}\left(\frac{\partial p}{\partial \tilde{T}}\right)_n=\rho+p-n\left(\frac{\partial
\rho}{\partial n}\right)_{\tilde{T}},
\end{equation}
Eq.~\eqref{78a} gives for the temperature evolution of a relativistic fluid in
the presence of particle creation the relation
\bea
\frac{\dot{\tilde{T}}}{\tilde{T}}&=&\left(\frac{\partial p}{\partial \rho}\right)_n\frac{\dot{n}}{n}=c_s^2\frac{\dot{n}}{n}=c_s^2\left(\Theta -3H\right)\nonumber\\
&&=-3Hc_s^2\left(1+\frac{P_c}{\rho +p}\right)=3Hc_s^2\left[\frac{\gamma^2}{\xi^2}\frac{u_{\mu}Q^{\mu}}{3H(\rho+p)}-1\right],\nonumber\\
\eea
where $c_s^2=\left(\partial p/\partial \rho \right)_n$ is the speed of sound in the newly created matter. If $\left(\partial p/\partial \rho\right)_n=c_s^2=\mathrm{
constant}$, for the temperature-particle number relation we obtain the
simple expression $\tilde{T} \sim n^{c_s^2}$.

\subsubsection{The case $w=-1$}

In the thermodynamical approach developed in the previous Sections we have assumed that particles are created in the form of ordinary baryonic matter, and therefore $w=p/\rho \geq 0$. Nevertheless, the open systems irreversible thermodynamic interpretation of the linear/scalar quadratic conformal $f\left(R,L_m\right)$ gravity  can be also extended to the case $w<0$. Next, we will investigate this problem, and we will show that our thermodynamical results,  in the sense of regularity and well-definiteness, are also valid even in the case of $w=-1$, that is, for matter satisfying the equation of state $\rho +p=0$ \cite{Wu1}.

We consider again the temperature evolution equation,
\begin{equation}\label{tempevol}
\frac{\dot{\tilde{T}}}{\tilde{T}}
=\left(\frac{\partial p}{\partial \rho}\right)_n\frac{\dot{n}}{n},
\end{equation}
and we will show that it is still valid even if $w= p/\rho = -1$. To establish this result,  we begin with the perfect-fluid energy-momentum balance equation
\begin{eqnarray}\label{61}
&&\dot{\rho}+3(\rho + p) H = \frac{\gamma^2}{\xi^2}u_{\mu}Q^{\mu}.
\end{eqnarray}
When $w=-1$, Eq.~\eqref{61} becomes
\begin{equation}
  \dot{\rho} = \frac{\gamma^2}{\xi^2}u_{\mu}Q^{\mu}\equiv -3H P_c.
\end{equation}
 By assuming adiabatic particle production, with $\dot{\sigma}=0$, the Gibbs law gives
\begin{equation}
\dot{\rho} = (\rho+P)\frac{\dot{n}}{n} = 0.
\end{equation}
 From the above two equations, we immediately obtain
\begin{equation}
  \dot{\rho} = P_c = 0.
\end{equation}
Since $\rho = \rho \left(n, \tilde{T}\right)$, we have
\begin{equation}
\dot{\rho}=\left(\frac{\partial \rho }{\partial n}\right)_{\tilde{T}} \dot{n} +\left(\frac{\partial \rho}{\partial \tilde{T}}\right)_n\dot{\tilde{T}}
= 0.
\end{equation}
By taking into account the thermodynamic identity (\ref{Termid}) for $\rho +p=0$, we obtain
\begin{equation}
\tilde{T}\left(\frac{\partial P}{\partial \tilde{T}}\right)_n
=  -n\left(\frac{\partial\rho}{\partial n}\right)_{\tilde{T}}.
\end{equation}

Hence it immediately follows that Eq.~\eqref{tempevol} with $\left(\partial p/\partial \rho\right)_n$ is valid even for $w=-1$, and generally for any negative values of $w$. For $w=-1$,  from Eq.~\eqref{tempevol} it follows that $n\tilde{T}$ is a constant, or $\tilde{T}\sim 1/n$. This relation indicates if the density of the "dark energy" particles is extremely low, their thermodynamic temperature is very high. If the dark energy particle number density is high, their temperature is very low. If $n\rightarrow \infty$, the temperature of the system made of dark energy particles tends to zero.

\section{Cosmological applications}\label{sect4}

In order to investigate the cosmological implications of the linear/scalar quadratic conformally invariant $f\left(R,L_m\right)$ theory, we consider that the Universe is isotropic and homogeneous, and that its geometry can be described by the Friedmann-Lemaitre-Robertson-Walker (FLRW) metric,
\be
ds^2=dt^2-a^2(t)\delta _{ij}dx^idx^j=a^2(\eta)\left(d\eta ^2-\delta _{ij}dx^idx^j\right),
\ee
where $a(t)$ is the dimensionless scale factor, and $\eta$ is the conformal time, defined as $dt=ad\eta$, or $\eta =\int{dt/a(t)}$. Moreover, we also introduce the Hubble function, defined as $H=\dot{a}/a=\left(1/a^2\right)\left(da/d\eta\right)$, where a dot denotes the derivative with respect to the cosmological time $t$. For the baryonic matter content of the Universe, we assume that it is represented by a perfect fluid, with matter Lagrangian $L_m=\rho$, and energy-momentum tensor with components given in the comoving frame by
\begin{align}
T^\mu_\nu={\rm diag}(\rho,-p,-p,-p),
\end{align}
where $\rho$ is the energy density of the cosmological matter, and $p$ is the pressure. In the following, we will consider the late time behavior of the model, and consequently we will assume that the Universe is filled by dust with equation of state of the form $p=0$.

\subsection{Cosmological equations of the Weyl vector}

In a cosmological geometry described by the flat Friedmann-Lemaitre-Robertson-Walker metric, the Weyl vector field equations take the form
\be
\frac{\partial}{\partial x^{\sigma}}\tilde{F}_{\mu \nu}+\frac{\partial}{\partial x^{\nu}}\tilde{F}_{\sigma \mu }+\frac{\partial}{\partial x^{\mu}}\tilde{F}_{ \nu \sigma}=0,
\ee
\be
\frac{1}{\sqrt{-g}}\frac{\partial }{\partial x^{\mu}}\left(\sqrt{-g}\tilde{F}^{\mu \nu}\right)+\frac{3}{%
	2}M_p^2\alpha ^{2}\delta^2\mathcal{L}_m \omega ^{\nu}\sqrt{-g}=0,
\ee
where
\begin{equation}
\tilde{F}_{\mu \nu }=a^{2}\left( \eta \right)
\begin{pmatrix}
0 & \tilde{E}_{1} & \tilde{E}_{2} & \tilde{E}_{3} \\
-\tilde{E}_{1} & 0 & -\tilde{B}_{3} & \tilde{B}_{2} \\
-\tilde{E}_{2} & \tilde{B}_{3} & 0 & -\tilde{B}_{1} \\
-\tilde{E}_{3} & -\tilde{B}_{2} & \tilde{B}_{1} & 0%
\end{pmatrix}%
\end{equation}%
with the vector fields $\left( \tilde{E}_{i},\tilde{B}_{i}\right) $ defined
in the ordinary Minkowski geometry.
We represent the Weyl vector as $\omega _{\mu}=\left(a^2\omega _0,a^2\vec{\omega}\right)$.
Hence we obtain the equations describing the cosmological evolution of the Weyl field as
\begin{align}\label{142}
-\frac{1}{a^{2}}\frac{\partial }{\partial \eta }(a^{2}\vec{\tilde{B}})%
+\nabla \times \vec{\tilde{E}}=0,
\end{align}
\begin{align}
\nabla \cdot \vec{\tilde{B}}=0,
\end{align}
\begin{equation}\label{143}
\nabla \times \vec{\tilde{B}}+\frac{1}{a^{2}}\frac{\partial }{\partial
\eta }a^{2}\vec{\tilde{E}}+\frac{3}{%
	2}M_p^2\alpha ^{2}\delta^2a^2\mathcal{L}_m \vec{\omega}=0,
\end{equation}
\begin{equation}\label{144}
\nabla \cdot \vec{\tilde{E}}-\frac{3}{%
	2}M_p^2\alpha ^{2}\delta^2a^2\mathcal{L}_m \omega_0=0.
\end{equation}

\subsection{The energy-momentum tensor of the Weyl field}

Since the FLRW geometry is isotropic, Weyl vector fields can
exist in such Universe only if one assumes that they have a random distribution, and an averaging procedure is performed. Thus, we suppose that the Weyl electric and magnetic fields satisfy the following conditions \cite{Nov1,Nov2,Nov3},
\begin{equation}
\left\langle \tilde{E}_{i}\right\rangle =0,\quad\left\langle \tilde{B}%
_{i}\right\rangle =0,\quad\left\langle \tilde{E}_{i}\tilde{B}_{j}\right\rangle =0,
\end{equation}
\begin{equation}
\left\langle \tilde{E}_{i}\tilde{E}_{j}\right\rangle =\frac{1}{3}%
\left\langle \vec{\tilde{E}}^{2}\right\rangle \delta_{ij},\quad\left\langle \tilde{B}_{i}%
\tilde{B}_{j}\right\rangle =\frac{1}{3}\left\langle \vec{\tilde{B}}%
^{2}\right\rangle \delta_{ij},
\end{equation}
where $\left\langle X\right\rangle$, representing the spatial average of a quantity $X$ on a given volume and at a fixed time, is defined as
\begin{align}\label{ave}
\left\langle X\right\rangle=\frac{1}{V_0}\lim_{V\rightarrow V_0}\int{\sqrt{-g}Xd^3x^i}.
\end{align}

Hence, the 00 component of the energy-momentum tensor of the Weyl field, $\tilde{T}^{(\omega)}_{00}=\left(a^2/4\delta ^2\right)\left(\vec{\tilde{E}}^2+\vec{\tilde{B}}^2\right)$ becomes
\be
\left\langle \tilde{T}^{(\omega )}_{00}\right\rangle =\frac{a^2}{4\delta ^2}\left(\left\langle \vec{\tilde{E}}^{2}\right\rangle+\left\langle \vec{\tilde{B}}^{2}\right\rangle\right),
\ee
while the diagonal components, given by
\be
\tilde{T}^{(\omega)}_{ik}=-\frac{a^2}{2\delta ^2}\left[\tilde{E}_i\tilde{E_k}+\tilde{B}_i\tilde{B_k}-(1/2)\left(\vec{\tilde{E}}^2+\vec{\tilde{B}}^2\right)\delta_{ik}\right],
\ee
are obtained as
\be
\left\langle \tilde{T}^{(\omega )}_{ik}\right\rangle=\frac{1}{3}\left\langle \tilde{T}^{(w)}_{00}\right\rangle \delta_{ik}.
\ee

Hence we have obtained the important result that the contribution of the Weyl vector to the cosmological dynamics can be modelled via a Weyl fluid, with effective energy density $\rho _{\omega}$ and effective pressure $p_{\omega}$, and energy-momentum tensor given by
\be
T_{\mu\nu}^{(\omega)}=\left(\rho_{\omega}+p_{\omega}\right)u_{\mu}u_{\nu}-p_{\omega}g_{\mu\nu},
\ee
where
\be
\rho_{\omega}=\frac{p_{\omega}}{3}, \rho _{\omega}=\frac{1}{4\delta ^2}\left(\left\langle \vec{\tilde{E}}^{2}\right\rangle+\left\langle \vec{\tilde{B}}^{2}\right\rangle\right).
\ee
Now, let us consider the covariant divergence of the vector field equation \eqref{Proca1}. Noting the antisymmetric nature of the tensor $\tilde{F}_{\mu\nu}$, one obtains
\begin{align}
\omega^\nu\nabla_\nu(L_m)=0,
\end{align}
where we have used the gauge condition $\nabla_\mu\omega^\mu=0$.
The matter Lagrangian of a perfect fluid is $L_m=\rho$, which depends only on $\eta$ in FRW universe. As a result, the above relation implies that on top of the FRW space-time, one has
\begin{align}
\omega_0=0.
\end{align}
In terms of the Weyl vector field $\omega_\mu$, the averaged cosmological electric and magnetic fields can be obtained as
\be
\vec{\tilde{B}}=\nabla\times\vec{\omega},
\ee
\be\label{164}
\vec{\tilde{E}}=\frac{1}{a^2}\frac{\partial}{\partial\eta}(a^2\vec{\omega}).
\ee

By assuming that the the fields $\vec{\omega}$, $\vec{\tilde{B}}$ and $\vec{\tilde{E}}$ are weak and very slowly in space, and also $\alpha\ll1$,  one can see that the magnetic field vanishes, and from \eqref{143} we obtain for the time variation of the Weyl electric field the relation
\be\label{165}
\vec{\tilde{E}}=\frac{\vec{\tilde{E}}_0}{a^2}.
\ee
where $\vec{\tilde{E}}_0$ is an integration constant.
Now, by taking the space average, we obtain
\be
\rho _{\omega}=\frac{1}{4\delta^2}\frac{\rho _0}{a^4},
\ee
where $\rho _{0}=\langle \vec{\tilde{E_0}}^2\rangle$ is a constant.

From Eqs.~(\ref{164}) and (\ref{165}) we immediately obtain
\be
a^2\vec{\omega}=\vec{E}_0\eta +\vec{E}_1,
\ee
where $\vec{E}_1$ is an integration vector. Since $\omega _{\mu}=a^2\left(\omega _0, \vec{\omega}\right)$, we obtain
\bea\label{168}
\left<\omega _i\omega _j\right>&=&a^4\left<\omega _i\omega _j\right>=\frac{1}{3}a^4\left<\vec{\omega}^2\right>\delta _{ij}\nonumber\\
&=&\frac{1}{3}\left<\left(\vec{E}_0\eta +\vec{E}_1\right)\cdot \left(\vec{E}_0\eta +\vec{E}_1\right)\right>\delta _{ij}\nonumber\\
&=&\frac{1}{3}\left[\left<\vec{E}_0^2\right>\eta ^2+2\left<\vec{E}_0\cdot \vec{E}_1\right>\eta+\left<\vec{E}_1^2\right>\right]\delta_{ij}\nonumber\\
&\equiv & X(\eta)\delta _{ij}.
\eea

\subsection{The generalized Friedmann equations}

In order to obtain the cosmological field equations for the FLRW metric (the generalized Friedmann equations), one should take an average of the metric field equations \eqref{feq}, by following the procedure outlined in Eq.~\eqref{ave}. As a result we obtain
\begin{align}\label{eq1}
(4H^2-2M_p^2\xi^2 a^2&-3\alpha^2X)(\xi^2\rho+\gamma^2)\nn
&+4\xi^2 H\dot{\rho}-\frac{2\gamma^2\rho_0}{3M_p^2\delta^2 a^2}=0,
\end{align}
and
\begin{align}\label{eq2}
(\xi^2\rho&-2\gamma^2)\dot{H}+(2\xi^2\rho-\gamma^2)\left(H^2-\frac34\alpha^2X\right)\nn
&-\xi^2H\dot\rho-\xi^2\ddot\rho+\frac32M_p^2\gamma^2\xi^2a^2-\frac{\gamma^2\rho_0}{6M_p^2\delta^2a^2}=0,
\end{align}
respectively. In obtaining the above equations, we have used the relation (\ref{168}), which can be reformulated equivalently as a differential equation,
with the function $X$ obtained as a solution of the second order ordinary differential equation,
\begin{align}\label{eq3}
\ddot{X}-\frac{2}{3}\rho_0=0.
\end{align}

We now define a set of dimensionless quantities
\begin{align}
&H_0\eta=\tau,\quad H=H_0h,\quad \bar{\rho}=\frac{\xi}{3M_p^2H_0^2}\rho,\nn
&\bar{\xi}=\frac{\xi^2H_0^2M_p^2}{\gamma^2},\quad\bar\rho_0=\frac{\alpha^2\bar{\delta}^2}{H_0^4}\rho_0,\quad\bar{\delta}=\frac{H_0}{\sqrt{6}M_p\alpha\delta},\nn
&\bar\gamma=\frac{\xi^2M_p^2}{2H_0^2},\quad \bar{X}=\frac{\alpha^2}{4H_0^4}X,
\end{align}
where $H_0$ is the current value of the Hubble parameter.
The field equations \eqref{eq1}, \eqref{eq2} and \eqref{eq3} are then simplified as
\begin{align}
(h^2-\bar\gamma a^2-3\bar{X})(1+3\bar\rho)+3h\dot{\bar{\rho}}-\frac{\bar{\rho}_0}{a^2}=0,
\end{align}
\begin{align}
(2-3\bar\rho)&\dot{h}+(1-6\bar\rho)(h^2-3\bar{X})+3(h+1)\dot{\bar{\rho}}\nonumber\\&+\frac{\bar{\rho}_0-3 a^4\bar\gamma}{a^2}=0,
\end{align}
\begin{align}
\ddot{\bar{X}}-\frac{\bar{\rho}_0}{6\bar\delta^2}=0.
\end{align}

\subsection{Comparison with the observational data}.

In order to compare the model with observational data, we transform the above equations into the redshift coordinates defined as $1+z=1/a$, giving $d/dt=-(1+z)h(z)d/dz$. Noting that the dimensionless Hubble parameter $h(z)$ has the property $h(z=0)=1$, one can obtain the following constraint on the model parameters
$$\bar{X}(z=0)=\frac{1-\bar\gamma-\bar{\rho}_0+3\Omega_{m0}(1-\bar\gamma)}{3(1+3\Omega_{m0})}$$

We estimate the values of the model parameters $H_0$, $\bar\gamma$ and $\bar\rho_0$ and $\dot{\bar{X}}(z=0)\equiv\dot{\bar{X}}_0$ by using the observational data on the Hubble parameter in the redshift range $0<z<2$, as presented in \cite{hubble}. We fix the value of the current value of the matter density $\Omega_{m0}$ to be equal to its $\Lambda CDM$ value $\Omega_{m0}=0.305$.

Also, since the constant $\bar\delta$ appears only in the evolution equation \eqref{eq3} of the function $\bar{X}$ as a modulator of the constant $\bar\rho_0$, we assume a fixed value $\bar\delta=0.1$ in the following calculations.

We use the likelihood analysis of the model based on the data on $H_0$.  The likelihood function in the case of $n$ independent data points can be defined as
\begin{align}
L=L_0e^{-\chi^2/2},
\end{align}
where $L_0$ is the normalization constant, and $\chi^2$ is defined as
\begin{align}
\chi^2=\sum_i\left(\frac{O_i-T_i}{\sigma_i}\right)^2,
\end{align}
where $i$ indicates the number of data, $O_i$ are the observational values, $T_i$ are the theoretical values and $\sigma_i$ are the observational errors associated with the $i$th data.
In the present model we have
\begin{align}
L=L_0\,\textmd{exp}\left[-\frac12\sum_i\left(\frac{O_i-H_0T_i}{\sigma_i}\right)^2\right],
\end{align}

By maximizing the likelihood function, one can find the best fit values of the parameters. In Table \ref{tab1}, we have summarized the result of the Maximum likelihood analysis together with their $1\sigma$ and $2\sigma$ confidence interval.
\begin{table}[h]
	\begin{center}
		\begin{tabular}{|c||c|c|c|}
			\hline
			Parameter&Best fit value&$1\sigma$~interval&$2\sigma$~interval\\
			\hline
			$H_0$&$67.47$&$67.47\pm1.41$&$67.47\pm2.76$\\
			\hline
			$\bar\gamma$&$0.012$&$0.012\pm0.05$&$0.012\pm0.10$\\
			\hline
			$\bar\rho_0$&$0.026$&$0.026\pm0.002$&$0.026\pm0.005$\\
			\hline
				$\dot{\bar{X}}_0$&$-0.199$&$-0.199\pm0.011$&$-0.199\pm0.022$\\
			\hline
		\end{tabular}
	\end{center}
	\caption{Best fit values of the linear model parameters $H_0$, $\bar\gamma$, $\bar\rho_0$ and $\dot{\bar{X}}_0$ together with their $1\sigma$ and $2\sigma$ confidence intervals.\label{tab1}}
\end{table}

\begin{figure*}[ht]
	\centering
	\includegraphics[scale=0.65]{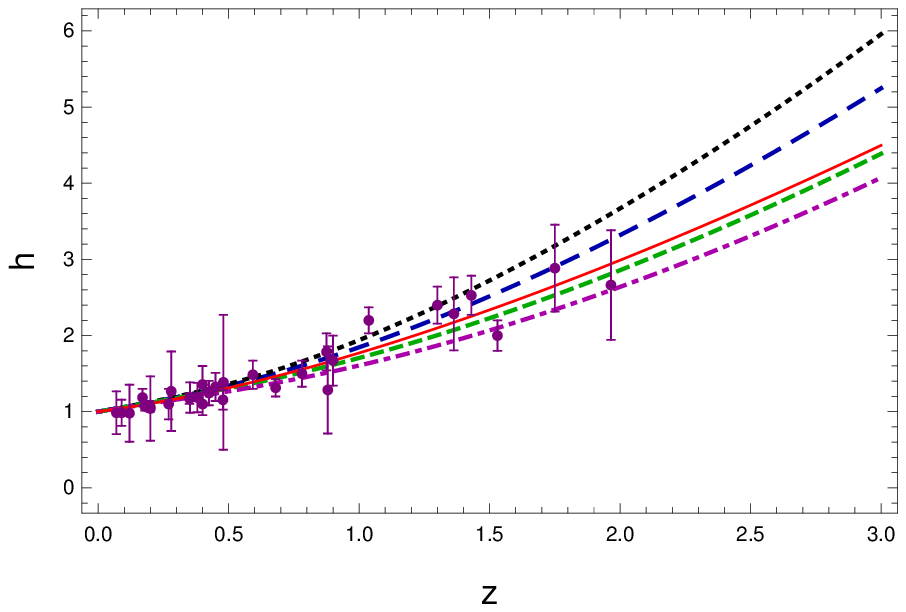}~~\includegraphics[scale=0.65]{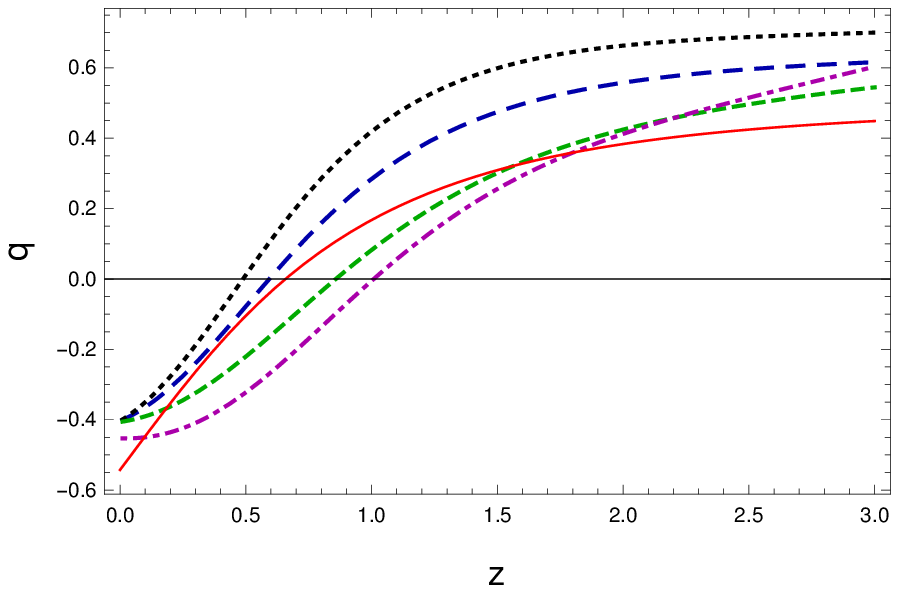}
	\caption{The evolution of the Hubble function $(1+z)h$  (left panel), and of the deceleration parameter $q$ (right panel)  as a function of redshift for the best fit values (long-dashed line), for $\bar\gamma=0.1$ and best fit values for $\bar\rho_0$ and $\dot{\bar{X}}_0$ (dotted line), for  $\bar\rho_0=0.021$ and best fit values for $\bar\gamma$ and $\dot{\bar{X}}_0$
 (dashed line) and for $\dot{\bar{X}}_0=-0.22$, $\bar\rho_0=0.021$ and best fit value for $\bar\gamma$  (dot-dashed line) respectively. The solid red line corresponds to the $\Lambda$CDM model.  The error bars indicate the observational values \cite{hubble}.}\label{fig1}
\end{figure*}

\begin{figure*}[ht]
	\centering
	\includegraphics[scale=0.65]{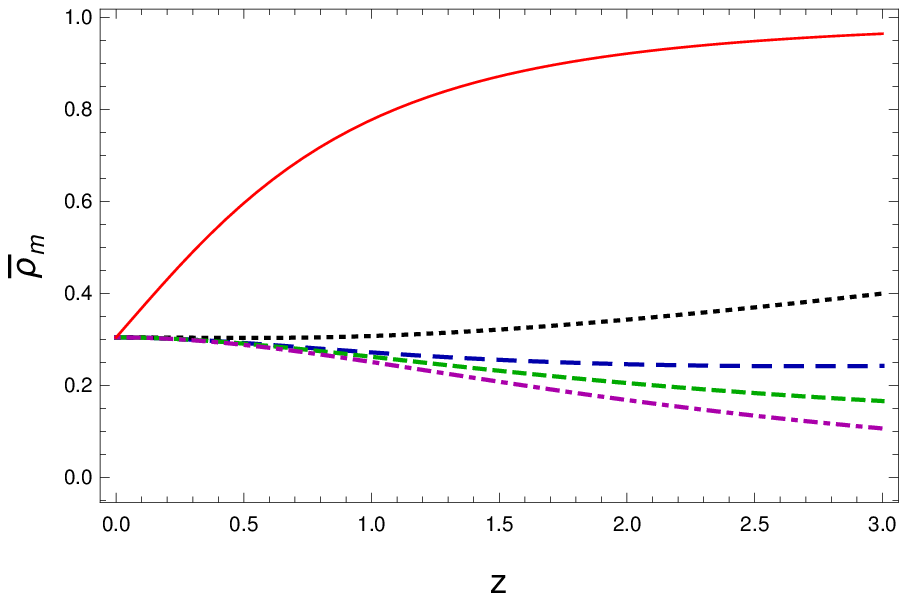}~~\includegraphics[scale=0.65]{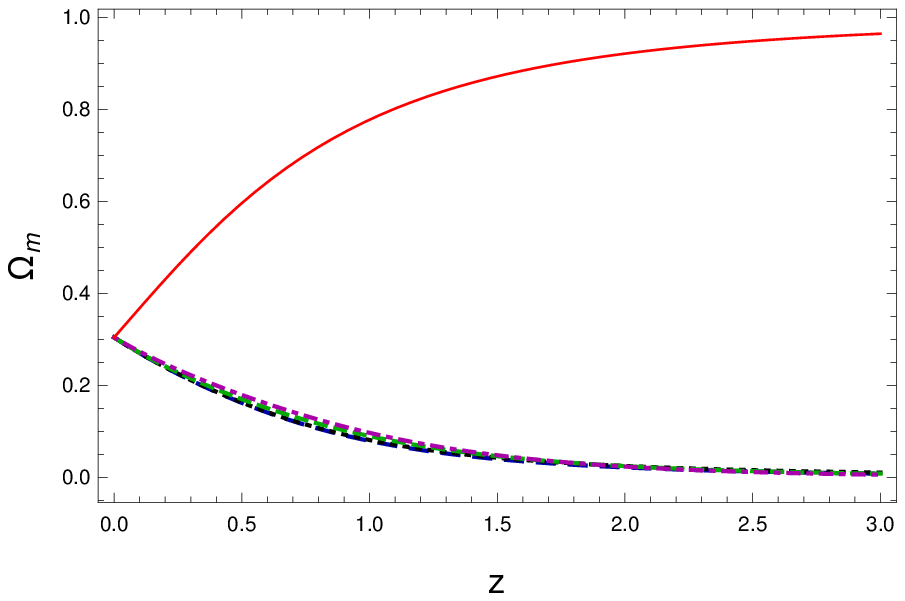}
	\caption{The evolution of the matter density $ \bar\rho_m$ (left panel) and of the matter density parameter $\Omega _m$ (right panel) as a function of redshift for the best fit values (long-dashed line), for $\bar\gamma=0.1$ and best fit values for $\bar\rho_0$ and $\dot{\bar{X}}_0$ (dotted line), for  $\bar\rho_0=0.021$ and best fit values for $\bar\gamma$ and $\dot{\bar{X}}_0$
		(dashed line) and for $\dot{\bar{X}}_0=-0.22$, $\bar\rho_0=0.021$ and best fit value for $\bar\gamma$  (dot-dashed line) respectively. The solid red line corresponds to the $\Lambda$CDM model.}\label{fig2}
\end{figure*}

The deceleration parameter and  the matter density are given by
\begin{align}
q=(1+z)\frac{h^\prime}{h},\quad \Omega_m=\frac{\bar\rho_m}{h^2(1+z)^2},
\end{align}
where the prime denotes the derivative with respect to the redshift. We have plotted the evolution of the Hubble parameter, of the deceleration parameter, and of the matter density as a function of redshift in Figures~\ref{fig1} and \ref{fig2}. One can see that the model could satisfy the observational data for the Hubble function very well, and it can reproduces almost exactly the predictions of $\Lambda$CDM model for $h(z)$ up to a redshift of $z\approx 3$. At a qualitative level the behavior of the deceleration parameter of the $\Lambda$CDM model is also recovered from the model. However, there are major differences in the behaviors of the matter densities, and of the matter density parameter, which show significant quantitative and qualitative differences as compared to $\Lambda$CDM. One possible explanation for the difference is that in the present model matter is not simply represented by $\rho$, but it has an effective meaning, with the contributions from the Weyl field and geometry-curvature coupling also giving some contributions to the "matter" content of the Universe.


\section{Discussions and final remarks}\label{sect5}

In formulating the first geometric theory of gravity Einstein extensively used the already known results of metric Riemannian geometry, in which the variation of the angles in the initial and final state of a vector rotated on a closed path is given by the curvature tensor. From a purely geometric point of view several extensions of Riemannian geometry can be considered, which lead to the introduction of new geometrical objects. One of these new objects is the torsion \cite{Hehl}, giving the non-closure of a parallelogram formed when two vectors are transported along each other. Finally, Weyl \cite{Weyl} considered geometries in which the variation of the length of a vector during parallel transport gives rise to the nonmetricity of the space-time. From a physical point of view the main goal of Weyl's approach was the geometric unification of gravity and electromagnetism. Einstein strongly criticized Weyl's {\it physical theory}, and this criticism led to the abandonment of the unified field theory approach proposed by Weyl. However, in  1929 Weyl \cite{Wg1,Wg2} showed that electrodynamics is invariant under the gauge transformations of the gauge field, and of the wave function of the charged field. Hence gauge theory, fundamental for particle physics, was born from Weyl's geometry. Another fundamental idea initially discussed by Weyl is the concept of conformal invariance. This is a highly attractive idea,  similar to the gauge principle in elementary particle physics that enriched so much contemporary physics. Global units transformations are analogous to global gauge transformations or global internal-symmetry transformations. The extension of units transformations to the local level, and the requirement of conformal invariance of physical laws is similar to the promotion of gauge and internal invariances to the local level by the introduction of gauge fields. Maxwell's equations, the massless Dirac equation, the massless scalar field equations, the electromagnetic, weak, and strong interactions between elementary particle fields are all conformally invariant. {\it Therefore, microscopic physics is conformally invariant in its entirety.} {\it However, Einstein's gravity is not}.

Hence, we do have another fundamental difference between the (geometric) world of particle physic/s, and the (geometric) world of the gravitational interaction. Since abandoning the conformal invariance of elementary particle physics is at least problematic, a possible bridge between quantum field theory and gravity can be constructed by imposing the principle of conformal invariance in Einstein gravity. This approach inevitably leads to the necessity of the extensive use of Weyl geometry to model gravitational phenomena.

In the present paper we have considered one of the simplest conformally invariant Weyl geometric models, initially introduced and studied from an elementary particle physics perspective in \cite{Gh1,Gh2,Gh3}, by including a new feature in the gravitational action, namely, the ordinary matter content. We have assumed a matter Lagrangian $L_m$ given in terms of the basic thermodynamic parameters of matter, the energy, density, or the pressure, respectively. However, in order {\it to build a conformally invariant gravitational theory in presence of matter a coupling between matter and curvature is necessary}. To maintain the conformal invariance in Weyl geometry of the gravitational theory we have adopted the simplest possible matter-geometry coupling, expressed by a term of the form $L_m\tilde{R}^2$. Hence the present theory is quadratic in the Weyl scalar $\tilde{R}$. The mathematical formalism can be significantly simplified by using the linear/scalar representation of quadratic Weyl gravity \cite{Gh1,Gh2,Gh3}, which allows to formulate the theory in Riemann geometry as a particular version of the already considered $f\left(R,L_m\right)$ type theories \cite{Lm,H4}, in which the gravitational action is represented as an arbitrary function of the (Riemannian) Ricci scalar, and of the matter Lagrangian. The theory introduced in the present paper imposes a specific restriction on the theory, namely, the requirement of conformal invariance.

After introducing the action of the theory in Weyl geometry, its representation as a linear/scalar model in Riemannian geometry was obtained. The field equations of the theory have been derived, and their various consequences have been discussed. In particular, it turns out that the divergence of the matter energy-momentum tensor does not generally vanish. Hence, this interesting property of the $f\left(R,L_m\right)$ gravity theories does also appear in their conformally invariant versions. We have briefly considered a thermodynamical interpretation of this effect, in terms of the irreversible particle creation by the gravitational field. Such an interpretation also imposes some strong constraints on the non-conservation vector $Q_{\nu}$. We have also considered the Newtonian limit of the theory, and obtained the generalized Poisson equation Eq.~(\ref{genPo}). In the linear approximation, as compared to the standard Poisson equation, two new terms does appear in the vacuum equation. The first one is proportional to the Newtonian potential itself, while the second is a new (free) term containing the Weyl vector, and the (effective) cosmological constant.

We have investigated the effects of the free term $3\left(\alpha \omega ^2/2+\xi ^2M_p^2\right)$ at the Solar System level, under the simplifying assumption $\omega ^2={\rm constant}$ by considering the problem of the perihelion precession of Mercury. By attributing the (very small) differences between the predictions of general relativity and observations to the presence of Weyl geometric effects a Solar System constraint on the product $\alpha ^2 \omega ^2/2$ can be obtained. Another set of constraints on the same quantities was obtained in \cite{Gh8} from a cosmological approach, giving $\alpha ^2\omega _3^2(0)\approx 0.22H_0^2$ and $\xi ^2\left(d\omega _3/dz\right)^2|_{z=0}\approx 1.24H_0^2$, where $H_0$ is the present value of the Hubble function.

It is interesting to note that keeping the term proportional to the potential in the generalized Poisson equation drastically modifies the potential. By denoting $\sigma =6\left(\xi ^2M_p^2-\alpha ^2\omega ^2/2\right)$, and $\chi =3\left(\alpha \omega ^2/2+\xi ^2M_p^2\right)$, respectively, in spherical symmetry Eq.~(\ref{Poieq1}) can be written as
\be
\frac{1}{r}\frac{d^2}{dr^2}\left(r\varphi\right)-\sigma \varphi -\chi=0.
\ee
This form of the Poisson equation is valid values of $r$ that do not satisfy the constraint (\ref{condnew}), that is, for values of the radial coordinate closer to the event horizon of the compact object. By assuming that the Weyl vector is constant, the general solution of the generalized Poisson equation is given by
\be
\varphi (r)=-\frac{\chi}{\sigma}+C_1\frac{e^{\sqrt{\sigma}r}}{r}+C_2\frac{e^{-\sqrt{\sigma}r}}{r},
\ee
where $C_1$ and $C_2$ are arbitrary constants of integration. If the condition $\sigma >0$, or, equivalently,  $\xi ^2M_p^2>\alpha ^2\omega ^2/2$, is satisfied, we obtain a Yukawa type gravitational potential $\varphi(r)=C_2\frac{e^{-\sqrt{\sigma}r}}{r}$, induced by the presence of the Weyl geometric effects. However, since $\xi ^2M_p^2$ can be interpreted as a cosmological constant, the effects of the exponential correction to the gravitational potential are negligibly small, at least at the level of the Solar System.

As we have already mentioned, and discussed in detail,  in the conformally invariant $f\left(R,L_m\right)$ theory, the ordinary matter energy-momentum tensor $T_{\mu \nu}$ is not conserved, and generally $\nabla _{\mu}T^{\mu \nu}\neq 0$. Conservative models in which the matter energy-momentum tensor is conserved, $\nabla _{\mu}T^{\mu \nu}=0$, can be obtained by imposing the condition $Q_{\nu}=0$, which would give a strong constraint on the Weyl vector $\omega _{\mu}$, which could be obtained in terms of the Ricci scalar and of the thermodynamical properties of the ordinary matter. On the other hand, the possible non-conservation of $T_{\mu \nu}$ has deep physical and astrophysical implications. One of its important consequences is the non-geodesic nature of the motion of free particles in a gravitational field, with the dynamical evolution taking place in the presence of an extra-force induced by the conformally invariant matter-curvature coupling.

In the Newtonian limit, the total acceleration $\vec{a}$ of an object moving in a gravitational field can be written as $\vec{a}=\vec{a}_N+\vec{a}_E$, where $\vec{a}_N$ is the usual Newtonian gravitational acceleration,  given by $\vec{a}_N=-GM\vec{r}/r^3$, where $M$ is the mass creating the field,  and $\vec{a}_E$ is the extra-acceleration. The acceleration equation immediately gives
$\vec{a}_E\cdot \vec{a}_N=\left(\vec{a}^2-\vec{a}_N^2-\vec{a}_E^2\right)/2$, and $ \vec{a}_N=\left(\vec{a}^2-\vec{a}_N^2-\vec{a}_E^2\right)\left[\vec{a}/\left(2\vec{a}_E\cdot \vec{a}\right)\right]+\vec{C}\times \vec{a}_E$,
respectively, where the arbitrary vector $\vec{C}$ can be taken as zero without any loss of generality. Finally, we can express the total acceleration as $\vec{a}=\tilde{a}_E\vec{a}_N$, where we have denoted
$1/\tilde{a}_E=(1/2)\left(\left|\vec{a}\right|/\left|\vec{a}_E\right|\right)\left(1-\vec{a}_N^2/\vec{a}^2-\vec{a}_E^2/\vec{a}^2\right)$.

Hence, it turns out that the total gravitational acceleration of a massive object moving in the field created by a mass $M$ is directly proportional to its Newtonian acceleration. Interestingly enough, a relation of this type, called the radial acceleration relation (RAR),  was found from the observations of the rotation curves of hydrogen clouds moving around the galactic center \cite{Rar1,Rar2,Rar3, Rar4}. The radial acceleration relation is an observational/empirical relation pointing towards the possible existence of a relationship between the centripetal acceleration $a_{obs} (r)= V_ {rot}^2 (r)/r$  due to the presence of dark matter in galaxies, and the  acceleration $a_{bar}(r)=V_{bar}^2/r$ of the baryonic matter. The RAR empirical relation is given by
\be
a_{obs}= f\left(\frac{a_{bar}}{a_+}\right)a_{bar},
\ee
where $f(x)$ is a fitting function to be determined from observations, and $a_+$ is a constant representing an acceleration scale. In the conformally invariant $f\left(R,L_m\right)$ the function $f\left(a_{bar}/a_+\right)$ corresponds to $\tilde{a}_E$, and thus may open some new possibilities for the observational test of the theory.

An important testing field of the conformally invariant $f\left(R,L_m\right)$ is cosmology. The presence of the Weyl vector induces an anisotropy in the cosmological expansion, and in order to maintain the isotropic and homogeneous nature of the Universe, an averaging procedure for the Weyl field is necessary. After taking spatial averaging it turns out that the energy-momentum tensor associated to the Weyl vector takes the form of a radiation fluid, satisfying an effective equation of state of the form $p=\rho/3$. A similar averaging procedure applied to the components of the Weyl vector leads to the system of generalized Friedmann equations  (\ref{eq1}) and (\ref{eq2}). The condition for the accelerated expansion has also been obtained. Depending on the numerical values of the model parameters, we obtain a large variety of  cosmological models. We have also performed a detailed comparison with the observational data, as well as with the predictions of the $\Lambda$CDM model. For specific values of the model parameters we find a good description of the observational data for $h(z)$, and a good concordance with the $\Lambda$CDM model at both low ($z<1$) and higher ($z\approx 3$) redshifts. The concordance is not so good for the deceleration parameter, and the matter density. However, the present investigations indicate that conformally invariant $f\left(R,L_m\right)$ gravitational models, theoretically consistent with the basic principles of the elementary particle physics, could lead to a novel understanding of the intricate dynamical properties of the Universe.

\section*{Acknowledgments}

We would like to thank the two anonymous reviewers for comments and suggestions that improved our work. The work of TH is supported by a grant of the Romanian Ministry of Education and Research, CNCS-UEFISCDI, project number PN-III-P4-ID-PCE-2020-2255 (PNCDI III).

\end{document}